\documentclass[11pt]{article}
\usepackage{amsmath}
\usepackage{graphicx}
\usepackage{color}
\usepackage{orcidlink}
\usepackage{hyperref}
\hypersetup{colorlinks=true, linkcolor=blue, citecolor=blue, urlcolor=blue}
\usepackage{caption}
\usepackage{subcaption}
\usepackage{setspace}
\usepackage{multirow}
\usepackage{multicol}
\usepackage{placeins}
\usepackage{amssymb}
\usepackage{float}
\usepackage{booktabs}
\RequirePackage{color}

\RequirePackage[numbers,sort&compress]{natbib}
\begin{document}
\baselineskip=20pt

\begin{center}
\setstretch{1.2}
\LARGE{Geodesic Structure, Thermodynamics and Scalar Perturbations of Mod(A)Max black hole Surrounded by Perfect Fluid Dark Matter }
\end{center}

\vspace{0.2cm}

\begin{center}
{\bf Faizuddin Ahmed\orcidlink{0000-0003-2196-9622}}\footnote{\bf faizuddinahmed15@gmail.com}\\
{\it Department of Physics, The Assam Royal Global University, Guwahati, 781035, Assam, India}\\

{\bf Ahmad Al-Badawi\orcidlink{0000-0002-3127-3453}}\footnote{\bf ahmadbadawi@ahu.edu.jo}\\
{\it Department of Physics, Al-Hussein Bin Talal University, 71111, Ma'an, Jordan}\\

{\bf Edilberto O. Silva\orcidlink{0000-0002-0297-5747}}\footnote{\bf edilberto.silva@ufma.br (Corresponding author)}\\
{\it Programa de P\'os-Gradua\c c\~ao em F\'{\i}sica \& Coordena\c c\~ao do Curso de F\'{\i}sica-Bacharelado, Universidade Federal do Maranh\~{a}o, 65085-580 S\~{a}o Lu\'{\i}s, Maranh\~{a}o, Brazil}
\end{center}

\vspace{0.4cm}

\begin{abstract}
\setstretch{1.4}
{\large In this work, we investigate the optical properties of a spherically symmetric Mod(A)Max black hole surrounded by perfect fluid dark matter, focusing on key features such as the photon sphere radius, shadow, photon trajectories, and the effective radial force experienced by photons. We also study the dynamics of massive particles around the black hole, deriving the effective potential and, from it, the specific energy and angular momentum of particles moving in circular orbits of fixed radii is discussed. The conditions for marginally stable circular orbits are analyzed, highlighting how the geometric parameters that modify the spacetime curvature influence both the optical and dynamical features. Furthermore, we explore the thermodynamic behavior of the black hole by examining its temperature, Gibbs free energy, and heat capacity, as well as its thermodynamic topology. Finally, scalar field perturbations are considered through the massless Klein-Gordon equation, and the quasinormal modes (QNMs) in the eikonal regime are computed, illustrating how the geometric parameters affect the potential and the QNM spectra. \\

 }
\end{abstract}

\pagebreak

\tableofcontents

\section{Introduction}

Black holes provide a uniquely sharp arena where strong-field gravity, high-energy field dynamics, and observational probes meet. In parallel with the classical pillars of black-hole physics and thermodynamics, including the area law and generalized second law \cite{Bekenstein1973,Bekenstein1974,GibbonsHawking1977,York1986}, Hawking radiation \cite{Hawking1975}, and the four laws of black-hole mechanics \cite{BardeenCarterHawking1973}, recent years have witnessed a rapid expansion of phenomenological frameworks aimed at modeling matter sectors beyond the Standard Model and quantifying their imprint on geodesic motion, lensing, shadows, quasinormal ringing, and thermodynamic response \cite{SC1984,Wald1984,Volker2022,SF1}. Among these frameworks, black holes embedded in effective dark-sector environments, such as perfect-fluid dark matter (PFDM), have attracted sustained attention as a tractable and observationally relevant way to encode dark matter effects in stationary spacetimes \cite{MH2012,HXZ2021,Kiselev2003}.

The PFDM paradigm is motivated by the fact that many astrophysical and cosmological observables are consistent with a dominant dark component while still leaving open its microscopic nature. A common approach is to incorporate dark matter through an effective energy-momentum tensor that preserves symmetries compatible with stationary black-hole geometries, allowing one to explore how a dark halo-like environment modifies the effective potential for photons and massive particles, the photon-sphere (light-ring) structure, and thus optical signatures such as strong deflection lensing and black-hole shadows \cite{SKJ2025,QQ2023,KJH2025} both in static and rotating black holes. This line of research has become especially timely after horizon-scale imaging by the Event Horizon Telescope (EHT) \cite{EHT2019_M87_I,EHT2019_M87_IV,EHT2019_M87_VI,EHT2019_M87_XII,EHT2019_M87_XIV,EHT2019_M87_XVI,EHT2019_M87_XVIi}, which elevates shadow observables from theoretical diagnostics to precision tools for testing gravity and matter sectors near compact objects.

Concurrently, nonlinear electrodynamics (NED) has long been recognized as a powerful mechanism to capture strong-field electromagnetic effects and, in some cases, to regularize curvature singularities in charged solutions. The historical prototype is Born-Infeld electrodynamics \cite{BornInfeld1934}, which inspired a large literature on black holes coupled to NED and related regular black-hole geometries \cite{AyonBeatoGarcia1998,AyonBeatoGarcia1999,Hayward2006}. Beyond its intrinsic theoretical appeal, NED is also phenomenologically motivated in regimes where standard Maxwell theory may be modified by vacuum polarization, high-field nonlinearities, or effective-field-theory operators \cite{HeisenbergEuler1936,DittrichGies2000}. In this landscape, ModMax electrodynamics stands out as a remarkable conformal and electromagnetic-duality-invariant nonlinear extension of Maxwell theory \cite{IB2020,BPK2020} (see also, \cite{MM1,MM2,MM3}). The ModMax theory has attracted considerable attention in this context. It introduces a dimensionless parameter $\gamma$ that controls the degree of nonlinearity of the electromagnetic field, smoothly interpolating between standard Maxwell theory and its nonlinear extension \cite{HMS2024}. The availability of exact and controlled deformations makes ModMax an attractive arena to revisit classical black-hole questions-geodesic structure, optical signatures, quasinormal spectra, and thermodynamic stability-in a setting where the matter sector is nontrivially modified while preserving key symmetries \cite{DFA2021,BES2026}. Some recent investigations have explored the thermodynamics, accretion dynamics, shadow images, perturbative properties, and scattering processes of ModMax black hole solutions in general relativity as well as modified gravity theories (see Refs.~\cite{ALBADAWI2025101865,ModMax1,ModMax2,ModMax3,ModMax4,ModMax5,ModMax6,ModMax7,ModMax8} and references therein).

In addition, recent developments consider ``phantom'' variants of electromagnetic sectors, where an effective sign flip in the electromagnetic contribution can mimic exotic components and lead to qualitatively new gravitational behavior. Such setups are frequently explored in the context of regular solutions, modified phase structure, and stability properties in asymptotically (A)dS spacetimes \cite{BES2026}. The interplay between a PFDM environment and (phantom) nonlinear electrodynamics is therefore a natural and timely direction: PFDM encodes external dark-sector dressing, while ModMax/Mod(A)Max encodes intrinsic strong-field modifications of the gauge sector. Their combined effects can reshape photon rings and effective potentials, alter the location and stability of circular orbits, and ultimately shift observable quantities such as shadow size and distortion, strong-lensing coefficients, and thermodynamic response functions \cite{Volker2022,SF1,SF2,SF3,SF4,VC2009}.

From the viewpoint of dynamics and strong-field probes, null geodesics and the associated photon-sphere (or more generally, light-ring) structure play a central role. Photon spheres govern critical impact parameters and set the gross features of shadow boundaries and strong-deflection images \cite{ClaudelVirbhadraEllis2001,VirbhadraEllis2000,Bozza2002,PerlickTsupko2015,Tsukamoto2017}. They are also closely related to eikonal quasinormal modes in many settings, thereby connecting optical and ringdown diagnostics \cite{ReggeWheeler1957,Zerilli1970,Teukolsky1973,KokkotasSchmidt1999,VC2009,SF1,SF3,SF2,SF4}. In rotating spacetimes, the shadow encodes both spacetime geometry and observer setup; practical shadow observables and degeneracies have been studied extensively \cite{HiokiMaeda2009,JohannsenPsaltis2010,Johannsen2013,BambiEtAl2017,CunhaHerdeiro2018,Gralla2019}. A complementary and increasingly important line of work studies strong-field lensing in a unified geometric-optics framework, including plasma or medium effects and generic observer-source configurations \cite{PerlickTsupko2015,TsupkoBisnovatyiKogan2015,Tsukamoto2017,Volker2022}. These methods provide the natural toolkit for assessing how PFDM and nonlinear gauge sectors jointly deform photon dynamics.

Thermodynamics provides the second major diagnostic channel. The semiclassical thermodynamics of black holes is foundational \cite{Bekenstein1973,Bekenstein1974,Hawking1975,BardeenCarterHawking1973,GibbonsHawking1977}, while modern developments emphasize stability and phase structure, particularly in the extended phase space where a cosmological constant acts as pressure \cite{HawkingPage1983,KubiznakMann2012,Dolan2014,Johnson2014}. In parallel, information-geometric methods (Weinhold and Ruppeiner geometries) have matured into practical probes of microscopic interactions and criticality in gravitational thermodynamics \cite{Weinhold1975,Ruppeiner1995}. These tools have been applied broadly to charged and rotating black holes and to diverse matter sectors, including NED-inspired models, to characterize local stability, heat-capacity transitions, and critical points \cite{Wei2022a,Wei2022b,Ruppeiner2008Review}. For PFDM-dressed black holes, thermodynamic response functions can acquire nontrivial parameter dependence, and the competition between PFDM dressing and nonlinear/phantom gauge contributions can generate qualitatively distinct stability windows \cite{HXZ2021,BES2026}. Closely related investigations combining exotic matter sectors (including phantom-like contributions), nonlinear electrodynamics, and dark components have also examined geodesic observables, shadows, perturbative potentials, greybody factors, and thermodynamic stability, providing complementary benchmarks for the present study \cite{FA1,FA2,FA6,FA7,FA8,FA11,FA12}.

Motivated by this confluence of observational and theoretical developments, in this work we investigate the geodesic structure and thermodynamic properties of a phantom ModMax or Mod(A)Max black hole immersed in PFDM. Noted that an ordinary ($\eta=+1$) or a phantom (($\eta=-1$)) ModMax black hole is characteristics by a constant $\eta=\pm\,1$ which is coupled with the ModMax parameter via $\eta\,e^{-\gamma}$. On the dynamical side, we analyze both null and timelike geodesics, characterize circular-orbit structure and the corresponding effective potentials, and extract shadow- and lensing-relevant quantities, with particular attention to how PFDM and the (phantom) nonlinear electrodynamics sector reshape the photon sphere and critical impact parameters \cite{VirbhadraEllis2000,Bozza2002,ClaudelVirbhadraEllis2001,PerlickTsupko2015,Volker2022,VC2009,SF1,SF2,SF3,SF4}. On the thermodynamic side, we compute the Hawking temperature, heat capacity, and related stability indicators, and we comment on how these results fit within the broader modern picture of black-hole thermodynamics and phase structure \cite{HawkingPage1983,KubiznakMann2012,Dolan2014,Johnson2014,Weinhold1975,Ruppeiner1995,Wei2022a,Wei2022b}. Our goal is to provide a unified account of optical/trajectory diagnostics and thermodynamic response for this PFDM--Mod(A)Max configuration, highlighting parameter regimes where observable imprints and stability properties are most pronounced.

\section{Spherically Symmetric Mod(A)Max BH Surrounded by PFDM}

In this paper, we study Mod(A)Max black hole surrounded by perfect fluid dark matter. The geodesic structure, thermodynamic properties and scalar perturbations will be explored and discuss the result showing the effects of the electric charge, Mod(A)Max's parameter and PFDM.

The action describing the coupling of Einstein gravity with the Mod(A)Max electrodynamic fields and matter field is given by
\begin{equation}
\mathcal{S} = \frac{1}{16\pi} \int_{\mathcal{M}} d^4x \, \sqrt{-g} \left(R - 4 \eta \, \mathcal{L} \right) + \mathcal{S}_M,
\end{equation}
where $g = \det(g_{\mu\nu})$ is the determinant of the metric tensor $g_{\mu\nu}$, $R$ is the Ricci scalar, and $\mathcal{S}_M$ represents the action of additional matter fields and we will consider the perfect fluid dark matter as field in our investigation.

In the action above, $\mathcal{L}$ denotes the ModMax Lagrangian, defined as \cite{IB2020,BPK2020}
\begin{equation}
\mathcal{L} = \mathcal{S} \cosh \gamma - \sqrt{\mathcal{S}^2 + \mathcal{P}^2} \, \sinh \gamma,
\end{equation}
where $\gamma$ is the dimensionless ModMax parameter. The quantities $\mathcal{S}$ and $\mathcal{P}$ are the electromagnetic scalar and pseudoscalar invariants, respectively, defined by
\begin{equation}
\mathcal{S} = \frac{\mathcal{F}}{4}, \qquad \mathcal{P} = \frac{\widetilde{\mathcal{F}}}{4},
\end{equation}
with $\mathcal{F} = F_{\mu\nu} F^{\mu\nu}$ and $\widetilde{\mathcal{F}} = F_{\mu\nu} \widetilde{F}^{\mu\nu}$. Here, $F_{\mu\nu} = \partial_\mu A_\nu - \partial_\nu A_\mu$ is the electromagnetic field tensor, $A_\mu$ is the gauge potential, and $\widetilde{F}^{\mu\nu} = \frac{1}{2} \epsilon^{\mu\nu\rho\lambda} F_{\rho\lambda}$ is its dual. By setting $\gamma = 0$, the Lagrangian reduces to the standard Maxwell form: $\mathcal{L} = \mathcal{F}/4$. 

To construct electrically charged black hole solutions, we consider $\mathcal{P} = 0$. Furthermore, we include a perfect fluid dark matter (PFDM) contribution as the matter field. Consequently, the generalized Einstein field equations in the presence of Mod(A)Max electrodynamics and PFDM read \cite{DFA2021,HXZ2021}:
\begin{equation}
G_{\mu\nu} = 8\pi \left( T_{\mu\nu} + T^{\rm PFDM}_{\mu\nu} \right),
\end{equation}
where $T_{\mu\nu}$ is the energy-momentum tensor associated with the Mod(A)Max field, and $T^{\rm PFDM}_{\mu\nu}$ corresponds to the PFDM. The Maxwell equations for the charged case take the form
\begin{equation}
\partial_\mu \left( \sqrt{-g} \, e^{-\gamma} F^{\mu\nu} \right) = 0.
\end{equation}

The energy-momentum for perfect fluid dark matter is given by \cite{HXZ2021}
\begin{equation}
T^{\text{PFDM}}_{\,\,\mu\nu} = \mathrm{diag}\left(-\mathcal{E}_{PFDM},\, P_{r\,PFDM},\, P_{\theta\,PFDM},\, P_{\phi\,PFDM}\right),
\label{tensor}
\end{equation}
where the components are
\begin{equation}
\mathcal{E}_{PFDM} = -P_{r\,PFDM}= -\frac{\lambda}{8\pi r^3},
\label{energy-density}
\end{equation}
and
\begin{equation}
P_{\theta\,PFDM}= -\frac{\lambda}{16\pi r^3}=P_{\phi\,PFDM}.
\label{pressures}
\end{equation}

Thereby, incorporating perfect fluid dark matter, a spherically symmetric Mod(A)Max black hole is described by the following line element \cite{BES2026}
\begin{equation}
    ds^2 = -f(r)\,dt^2 + \dfrac{dr^2}{f(r)} + r^2 (d\theta ^2 + \sin ^2{\theta }\,d\varphi ^2),\label{metric}
\end{equation}
where the lapse function is given by\footnote[4]{One may extend this study for dyonic Mod(A)Max black holes \cite{DFA2021} with PFDM \cite{MH2012} as well as other field (say Quintessence \cite{Kiselev2003}), where the lapse function becomes:\\
\[f(r) = 1-\frac{2 M}{r}+\eta\,e^{-\gamma}\,\frac{(Q^2_e+Q^2_m)}{r^2}+\frac{\lambda}{r}\ln\!\frac{r}{|\lambda|}-\frac{N}{r^{3 w+1}}\] with $Q_e$ and $Q_m$,respectively are the electric and magnetic charges.
}
\begin{align}
     f(r) = 1-\frac{2 M}{r}+\eta\,\frac{e^{-\gamma}\,Q^2}{r^2}+\frac{\lambda}{r}\ln\!\frac{r}{|\lambda|},\label{function}
\end{align}
Here $M$ represent mass of the black hole, $\lambda$ is the PFDM parameter, $\Lambda$ is the cosmological constant and $\eta=\pm\,1$ (plus sign corresponds to ModMax black hole and minus for phantom ModMax or Mod(A)Max black hole). It is worth noting here that Sekhmani \textit{et al.}~\cite{YS2025} studied the thermodynamic properties and phase transitions of anti-de Sitter black holes with ModMax nonlinear electrodynamics and perfect fluid dark matter, while Sulaman \textit{et al.}~\cite{SS2025} investigated the greybody factors and Hawking temperature of the same ModMax-AdS black hole configuration. Therefore, the current black hole configuration is differ from the ones already studied in the literature.

The space-time (\ref{metric}) can be expressed as $ds^2=g_{\mu\nu} dx^{\mu} dx^{\nu}$, where the metric tensor $g_{\mu\nu}$ is given by
\begin{equation}
    g_{\mu\nu}=\mbox{diag}\left(-f(r),\,\frac{1}{f(r)},\,r^2,\,r^2\, \sin^2 \theta\right),\quad (\mu,\nu=0,...,3).\label{bb1}
\end{equation}
The lapse function $f(r)$ at radial infinity behaves as,
\begin{equation}
    \lim_{ r \to \infty} f(r) \to 1\label{condition}
\end{equation}
which indicates the function is asymptotically flat.

\begin{figure}[tbhp]
    \centering
    \includegraphics[width=\textwidth]{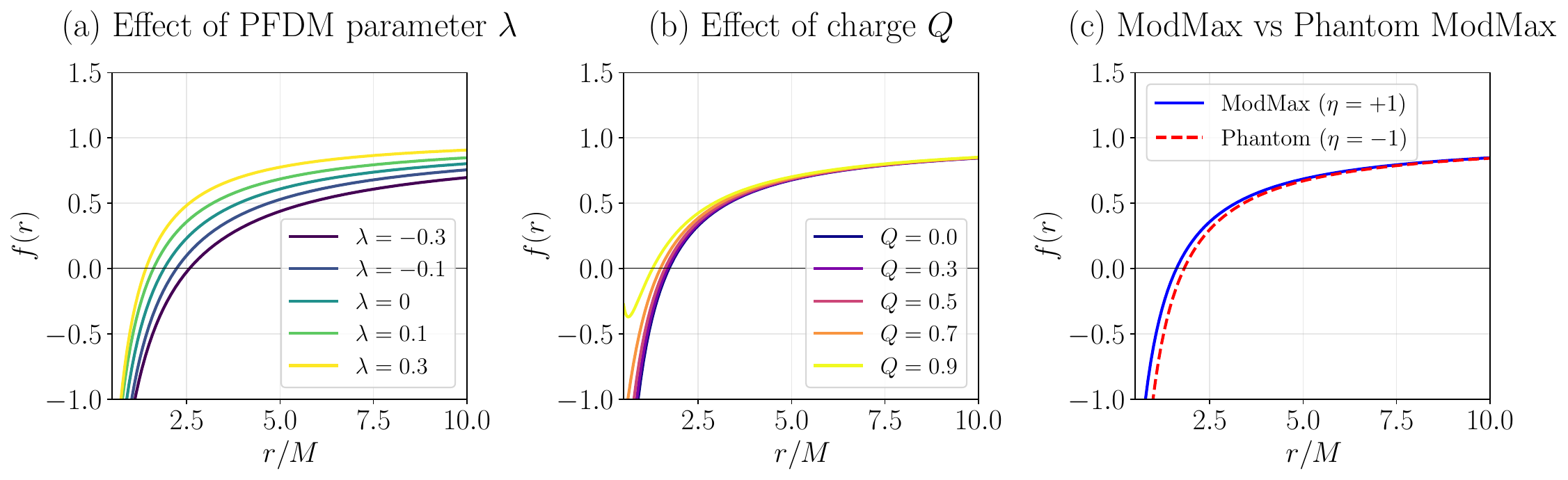}
    \caption{The lapse function $f(r)$ as a function of the radial coordinate $r/M$ for the Mod(A)Max black hole surrounded by perfect fluid dark matter. Panel (a) shows the effect of varying the PFDM parameter $\lambda$ while keeping $Q = 0.5M$ and $\gamma = 0.3$ fixed. Negative values of $\lambda$ increase the horizon radius, while positive values decrease it. Panel (b) illustrates the influence of the electric charge $Q$ on the metric function, demonstrating how larger charges lead to the emergence of inner and outer horizons characteristic of Reissner-Nordstr\"om-like solutions. Panel (c) compares the ModMax ($\eta = +1$) and phantom ModMax ($\eta = -1$) cases, revealing that the phantom configuration produces a larger event horizon due to the sign reversal in the electromagnetic contribution. The horizontal dashed line at $f(r) = 0$ marks the location of the event horizons.}
    \label{fig:lapse_function}
\end{figure}

The behavior of the lapse function $f(r)$ is crucial for understanding the causal structure of the spacetime. In Fig.~\ref{fig:lapse_function}, we present a detailed analysis of how the lapse function varies with the radial coordinate for different values of the model parameters. The zeros of $f(r)$ correspond to the event horizons of the black hole.

As shown in Fig.~\ref{fig:lapse_function}(a), the PFDM parameter $\lambda$ significantly affects the horizon structure. For $\lambda < 0$, the event horizon radius increases compared to the $\lambda = 0$ case, while positive values of $\lambda$ lead to a smaller horizon. This behavior can be understood from the logarithmic term in Eq.~(\ref{function}), which changes the large-$r$ approach to $f(r)\to 1$ only mildly (since $\ln r / r \to 0$), but can substantially shift the root structure at moderate radii. In practical terms, the intersection point of each curve with the line $f(r)=0$ is the horizon radius $r_h$; thus, Fig.~\ref{fig:lapse_function}(a) provides an immediate geometrical reading of how PFDM ``inflates'' or ``deflates'' the event horizon.

Panel (b) isolates the role of the electric charge. Increasing $Q$ enhances the $Q^2/r^2$ contribution and can generate an additional (inner) root of $f(r)$, i.e., a two-horizon structure reminiscent of Reissner--Nordstr\"om geometries. In the parameter range shown, one can identify whether the spacetime has a single horizon (one crossing) or two horizons (two crossings); approaching the extremal limit corresponds to the merger of these roots, visible as a tangency to $f(r)=0$.

Panel (c) highlights the effect of $\eta=\pm 1$, i.e., the sign choice of the nonlinear electromagnetic sector. For $\eta=+1$ (ModMax) the electromagnetic term behaves as in standard charged solutions, whereas for $\eta=-1$ (phantom Mod(A)Max) it effectively reverses its contribution in the metric function. The net consequence is a systematic shift of the horizon to larger radii in the phantom case, which directly impacts all subsequent observables that depend on the near-horizon geometry (photon sphere location, shadow size, and the structure of effective potentials).

\section{Photon Sphere and Shadow}

Null geodesics, corresponding to the trajectories of massless particles such as photons, play a fundamental role in understanding the causal structure of spacetime and the observable properties of black holes~\cite{SC1984,Wald1984}. In particular, they determine the location of the photon sphere, which is the spherical region where light can orbit the black hole in unstable circular orbits. The behavior of null geodesics governs the formation of black hole shadows, gravitational lensing patterns. Moreover, analyzing null geodesics allows one to derive effective potentials that capture the instability of circular orbits and identify critical impact parameters relevant for observational signatures. As a result, null geodesics provide a direct link between the theoretical geometry of black holes and measurable astrophysical phenomena such as the angular size of the shadow and the bending of light by strong gravitational fields.

The Lagrangian density function in terms of the metric tensor $g_{\mu\nu}$ is given by
\begin{equation}
    \mathbb{L}=\frac{1}{2}g_{\mu\nu} \dot x^{\mu}\,\dot x^{\nu},\label{bb2}
\end{equation}
where dot represent ordinary derivative w. r. to $\tau$, an affine parameter along geodesics.

On equatorial plane defined by $\theta=\pi/2$, the Lagrangian density function (\ref{bb2}) using the metric (\ref{metric}) explicitly written as
\begin{equation}
    \mathbb{L}=\frac{1}{2}\left[-f(r) \dot t^2+\frac{1}{f(r)} \dot r^2+r^2\,\dot \phi^2\right].\label{bb3}
\end{equation}

The Lagrangian density function is independent of temporal coordinate $t$ and angular coordinate $\theta$. Therefore, there exist at least two Killing vectors, namely, $\xi^{\mu}_t \equiv \partial_{(t)}$ corresponds to time translation symmetry and $\xi^{\mu}_{(\phi)} \equiv \partial_{\phi}$ corresponds to rotational symmetry. Hence, the conserved quantities associated with these are given by
\begin{equation}
    \mathrm{E}=-\xi^{\mu}_t\,u^{\nu}\,g_{\mu\nu}=-g_{tt} \dot t=f(r)\,\dot t,\label{bb4}
\end{equation}
And
\begin{equation}
    \mathrm{L}=\xi^{\mu}_{\phi}\,u^{\nu}\,g_{\mu\nu}=g_{\phi\phi} \dot \phi=r^2\,\dot \phi,\label{bb5}
\end{equation}
Here $\mathrm{E}$ is the conserved energy and $\mathrm{L}$ is the conserved angular momentum.

The radial equation of motion for photon particles can be expressed as
\begin{equation}
    \dot r^2+V_{\rm eff}=\mathrm{E}^2,\label{bb6}
\end{equation}
which is equivalent to the one-dimensional equation of motion of unit mass particles having energy $\mathrm{E}^2$. Here $V_{\rm eff}$ is the effective potential that governs the photon dynamics and is given by
\begin{equation}
    V_{\rm eff}=\frac{\mathrm{L}^2}{r^2}\,f(r)=\frac{\mathrm{L}^2}{r^2}\,\left(1-\frac{2 M}{r}+\eta\,\frac{e^{-\gamma}\,Q^2}{r^2}+\frac{\lambda}{r}\ln\!\frac{r}{|\lambda|}\right).\label{bb7}
\end{equation}

\begin{figure}[tbhp]
    \centering
    \includegraphics[width=\textwidth]{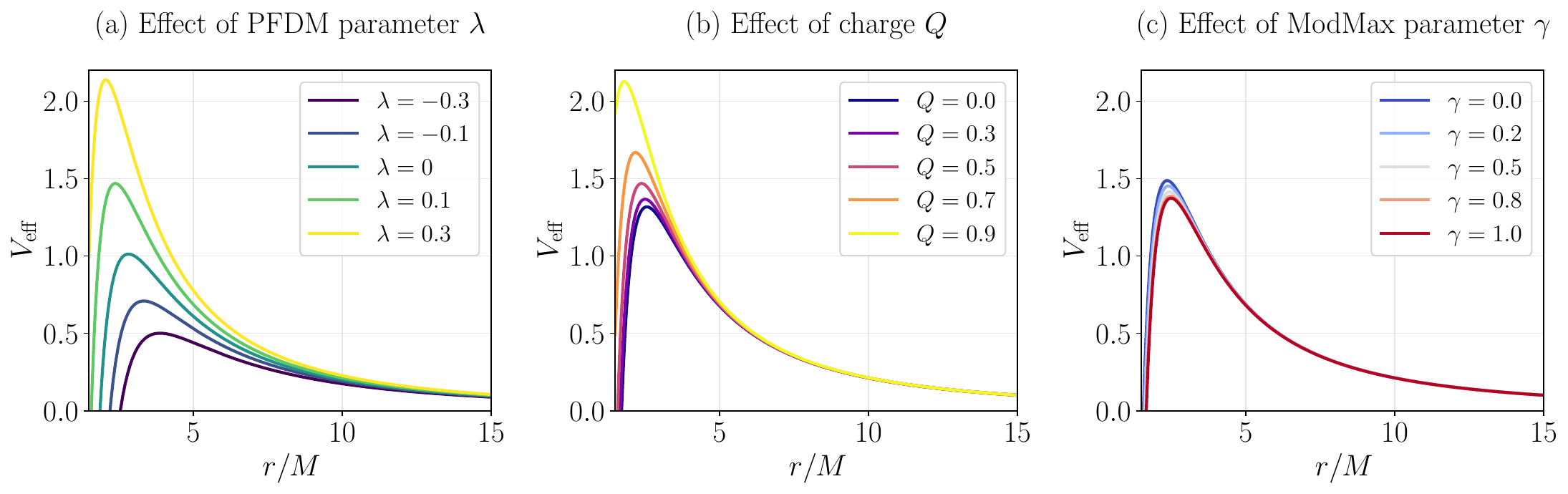}
    \caption{The effective potential $V_{\rm eff}$ for photon motion as a function of the radial coordinate $r/M$ in the Mod(A)Max black hole spacetime surrounded by perfect fluid dark matter. The angular momentum is set to $\mathrm{L} = 5M$. Panel (a) demonstrates the influence of the PFDM parameter $\lambda$: negative values of $\lambda$ increase the peak height and shift it outward, while positive values suppress the potential barrier. Panel (b) shows the effect of the electric charge $Q$, where increasing charge raises the potential barrier, indicating stronger photon deflection. Panel (c) illustrates how the ModMax parameter $\gamma$ modifies the potential; larger values of $\gamma$ reduce the electromagnetic contribution through the $e^{-\gamma}$ factor, lowering the potential peak. The maximum of $V_{\rm eff}$ corresponds to the unstable photon sphere, whose location and height directly influence the black hole shadow size.}
    \label{fig:photon_potential}
\end{figure}

The effective potential derived above depends explicitly on the electric charge $Q$, the PFDM parameter $\lambda$, the ModMax parameter \(\gamma\), and the black hole mass \(M\). Consequently, observable features such as the photon sphere (or photon ring), black hole shadow, and gravitational lensing are also affected by these parameters. In Fig.~\ref{fig:photon_potential}, we present a comprehensive analysis of the photon effective potential for various parameter choices.

Figure~\ref{fig:photon_potential} should be read as follows. For fixed $\mathrm{L}$, the potential $V_{\rm eff}(r)$ develops a single barrier outside the horizon; its maximum marks the unstable circular null orbit (photon sphere). Photons coming from infinity with a given impact parameter (equivalently, given ratio $\mathrm{L}/\mathrm{E}$) either (i) turn around at a radius where $\mathrm{E}^2=V_{\rm eff}$ (scattering), or (ii) cross the barrier and fall into the black hole (capture). Thus, the \emph{height} of the maximum sets the capture threshold, while the \emph{position} of the maximum sets the photon sphere radius $r_s$. 

Figure \ref{fig:photon_potential}(a) isolates the PFDM influence. Negative $\lambda$ increases the effective barrier and shifts it outward, which physically means that PFDM can move the light ring to larger radii and increase the critical impact parameter. Positive $\lambda$ does the opposite: it lowers the barrier and shifts the maximum inward, favoring capture and decreasing the shadow scale. This directly anticipates the monotonic trend later observed for the shadow radius.

Figure \ref{fig:photon_potential}(b) shows that increasing $Q$ typically strengthens the barrier (for fixed $\eta$), enhancing strong-field light deflection and shifting the unstable orbit. Figure \ref{fig:photon_potential}(c) clarifies the role of the ModMax parameter $\gamma$: because the electromagnetic sector enters as $e^{-\gamma}Q^2$, larger $\gamma$ suppresses the charge contribution in $f(r)$ and therefore reduces the electromagnetic imprint in $V_{\rm eff}$. In the limit of sufficiently large $\gamma$, the curves approach those of an effectively uncharged PFDM geometry, explaining why optical observables tend toward their Schwarzschild--PFDM counterparts.

The effective potential analysis reveals several important features. First, the maximum of $V_{\rm eff}$ corresponds to the location of the photon sphere, where circular null orbits exist but are unstable against radial perturbations. Photons with energy $\mathrm{E}^2$ less than the potential maximum will be reflected back to infinity (scattered photons), while those with energy exceeding this threshold will fall into the black hole (captured photons). The height and position of the potential barrier are thus directly related to the critical impact parameter that separates these two regimes.

For circular null orbits, the conditions $\frac{dr}{d\lambda}=0$ and $\frac{d^2r}{d\lambda^2}=0$ must be satisfied. From Eq. (\ref{bb6}), we find the following relations:
\begin{equation}
    \mathrm{E}^2=V_{\rm eff}=\frac{\mathrm{L}^2}{r^2}\,f(r).\label{bb8}
\end{equation}
And
\begin{equation}
    \frac{d}{dr}\left(\frac{f(r)}{r^2}\right)=0\Rightarrow 2\,r\,f(r)-r^2\,f'(r)=0.\label{bb9}
\end{equation}
To make the discussion of characteristic radii quantitative, we report in Table~\ref{tab:radii1}-\ref{tab:radii2} the numerical values of the outer horizon radius $r_h$, the photon-sphere radius $r_s$, and the shadow radius $R_{\rm sh}$ for representative parameter sets $(Q,\gamma,\lambda,\eta)$ adopted in our plots. This summary complements Figs.~\ref{fig:lapse_function} and the photon-sphere/ISCO figures by providing explicit benchmark numbers, allowing a direct comparison of how the PFDM parameter $\lambda$, the Mod(A)Max parameter $\gamma$, the charge $Q$, and the choice $\eta=\pm1$ shift the near-horizon geometry and the corresponding observable scales.

The relation (\ref{bb9}) gives us the photon sphere radius $r=r_s$ satisfying the following polynomial equation as,
\begin{align}
r^2_s-3 M r_s+2\eta e^{-\gamma}\,Q^2+\frac{\lambda\,r_s}{2 }\,\left(3\,\ln\!\frac{r_s}{|\lambda|}-1\right)= 0.\label{bb10}
\end{align}

The exact analytical solution of the above polynomial will give the photon sphere radius $r_s$. Noted that the exact analytical solution is quite a challenging due to the presence of logarithmic function. However, one can determine the numerical results by selecting suitable values of parameters. Thereby, it becomes evident that the photon sphere depends on the electric charge $Q$, the PFDM parameter $\lambda$, and the ModMax parameter \(\gamma\). 

As we have checked that the lapse function asymptotically flat. Therefore, for a static distant observer, the shadow radius is given by \cite{Volker2022}
\begin{equation}
   R_{\rm sh}=\frac{r_s}{\sqrt{f(r_s)}}=\frac{r^2_s}{\sqrt{r^2_s-2 M r_s+\eta e^{-\gamma} Q^2+\lambda r_s\ln\!\frac{r_s}{|\lambda|}}}.\label{bb12}
\end{equation}
\begin{itemize}
    \item For $\eta=+1$ corresponding to the ModMax black hole with PFDM, the shadow radius simplifies as,
    \begin{equation}
   R_{\rm sh}=\frac{r^2_s}{\sqrt{r^2_s-2 M r_s+e^{-\gamma} Q^2+\lambda r_s\ln\!\frac{r_s}{|\lambda|}}}.\label{bb13}
\end{equation}

\item For $\eta=-1$ corresponding to the Mod(A)Max black hole with PFDM, the shadow radius simplifies as,
\begin{equation}
   R_{\rm sh}=\frac{r^2_s}{\sqrt{r^2_s-2 M r_s-e^{-\gamma} Q^2+\lambda r_s\ln\!\frac{r_s}{|\lambda|}}}.\label{bb14}
\end{equation}
\end{itemize}

\begin{figure}[ht!]
    \centering
    \includegraphics[width=\textwidth]{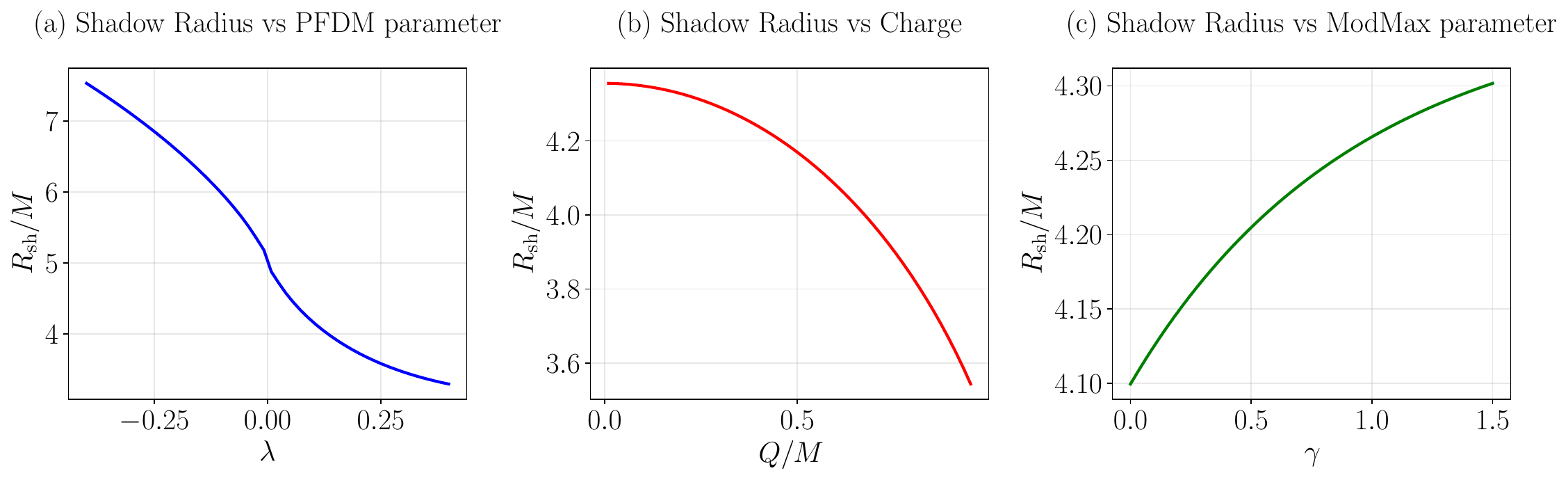}
    \caption{The shadow radius $R_{\rm sh}$ as a function of the model parameters for the Mod(A)Max black hole surrounded by perfect fluid dark matter. Panel (a) shows $R_{\rm sh}$ versus the PFDM parameter $\lambda$, revealing a monotonically increasing relationship: more negative (positive) values of $\lambda$ lead to larger (smaller) shadow radii. Panel (b) displays the shadow radius as a function of the electric charge $Q/M$, demonstrating that the shadow size decreases with increasing charge due to the enhanced electromagnetic repulsion that pushes the photon sphere inward. Panel (c) illustrates the dependence on the ModMax parameter $\gamma$; as $\gamma$ increases, the electromagnetic contribution is suppressed by the factor $e^{-\gamma}$, causing the shadow radius to approach that of a Schwarzschild black hole with PFDM. These results provide testable predictions for comparing with EHT observations of supermassive black hole shadows.}
    \label{fig:shadow_radius}
\end{figure}

The dependence of the shadow radius on the model parameters is illustrated in Fig.~\ref{fig:shadow_radius}. These results are particularly relevant for comparing theoretical predictions with observational data from the Event Horizon Telescope.

Figure~\ref{fig:shadow_radius} quantifies how the optical size of the black hole responds to each physical ingredient of the model through Eq.~(\ref{bb12}). In Fig. \ref{fig:shadow_radius}(a), varying $\lambda$ changes both the photon sphere radius $r_s$ (solution of Eq.~(\ref{bb10})) and the redshift factor $\sqrt{f(r_s)}$ entering $R_{\rm sh}$. The net result is monotonic over the range shown: negative $\lambda$ yields a larger $r_s$ and/or a smaller $\sqrt{f(r_s)}$, increasing $R_{\rm sh}$, while positive $\lambda$ reduces the shadow size. Figure \ref{fig:shadow_radius}(b) shows that increasing $Q$ decreases $R_{\rm sh}$ for the parameter domain considered. Operationally, a larger charge modifies $f(r)$ and typically shifts the unstable orbit inward, thereby reducing $r_s$ and the corresponding impact parameter of photons that asymptotically graze the photon sphere. Figure \ref{fig:shadow_radius}(c) demonstrates that $\gamma$ controls how effectively the charge contributes to the geometry: as $\gamma$ grows, the factor $e^{-\gamma}$ suppresses electromagnetic effects, so the shadow radius approaches that of a PFDM-deformed Schwarzschild spacetime. This provides a clear observational pathway: measuring a shadow size consistent with a weak charge imprint could be interpreted as either small $Q$ or sufficiently large $\gamma$ (degeneracy that can be lifted by combining with additional observables, e.g., lensing or accretion signatures).

\begin{figure}[ht!]
    \centering
    \includegraphics[width=\linewidth]{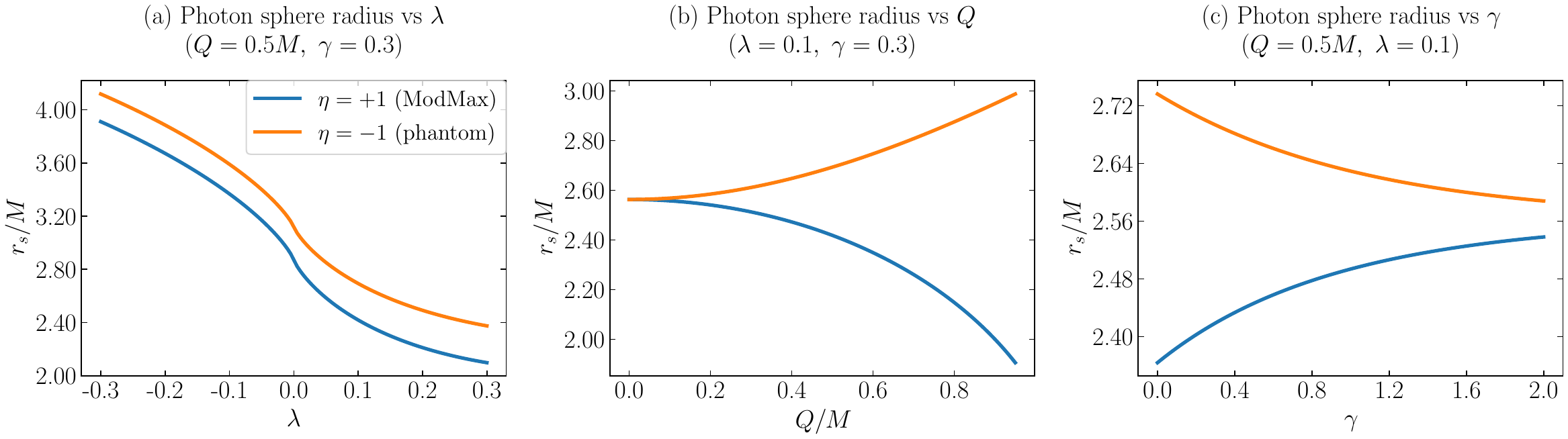}
    \caption{Photon-sphere radius $r_s$ (in units of $M$) as a function of the deformation parameter $\lambda$ and of the model couplings, for both branches $\eta=\pm1$. Panel (a) shows $r_s/M$ versus $\lambda$ at fixed $(Q,\gamma)$; panel (b) shows $r_s/M$ versus the charge-to-mass ratio $Q/M$ at fixed $(\lambda,\gamma)$; panel (c) shows $r_s/M$ versus $\gamma$ at fixed $(Q,\lambda)$. In all cases, the photon-sphere radius is obtained from the null-circular-orbit condition $2f(r_s)=r_s f'(r_s)$, selecting the physically relevant solution outside the event horizon.}
    \label{fig:rs_photonsphere}
\end{figure}
Figure~\ref{fig:rs_photonsphere} summarizes how the location of the photon sphere responds to the model deformation parameter $\lambda$ and to the couplings that control the effective charge sector. For each parameter sweep, we compute the outer horizon radius $r_h$ from $f(r_h)=0$ and then solve the null circular-orbit condition $2f(r_s)=r_s f'(r_s)$, retaining the root with $r_s>r_h$. Figure \ref{fig:rs_photonsphere}(a) isolates the direct impact of $\lambda$ on the unstable photon orbit for fixed $(Q,\gamma)$, while Figs. \ref{fig:rs_photonsphere}(b) and \ref{fig:rs_photonsphere}(c) display how the same orbit shifts under variations of the charge-to-mass ratio $Q/M$ and of the coupling $\gamma$, respectively, keeping the remaining parameters fixed. The comparison between the two branches $\eta=+1$ and $\eta=-1$ shows that the sign choice in the effective charge contribution leads to distinct trends for $r_s$, highlighting the sensitivity of the strong-lensing scale to the underlying branch structure of the theory.

Moreover, the critical impact parameter for photon particles at radius $r=r_s$ using Eq.(\ref{bb8}) is given by
\begin{equation}
    b_c=\frac{\mathrm{L}}{\mathrm{E}}\Big{|}_{r=r_s}=\frac{r^2_s}{\sqrt{r^2_s-2 M r_s+\eta e^{-\gamma} Q^2+\lambda r_s\ln\!\frac{r_s}{|\lambda|}}}=R_{\rm sh}.\label{bb15}
\end{equation}

Next, we determine the photon trajectories showing the effects of various parameters involved in the space-time metric. The equation of orbit using Eqs.~(\ref{bb5}) and (\ref{bb6}) is given by
\begin{equation}
    \left(\frac{dr}{d\phi}\right)^2=\frac{\dot r^2}{\dot \phi^2}=\frac{r^4}{b^2}-r^2 \left(1-\frac{2 M}{r}+\eta\,\frac{e^{-\gamma}\,Q^2}{r^2}+\frac{\lambda}{r}\ln\!\frac{r}{|\lambda|}\right).\label{bb16}
\end{equation}

\begin{figure}[ht!]
    \centering
    \includegraphics[scale=0.5]{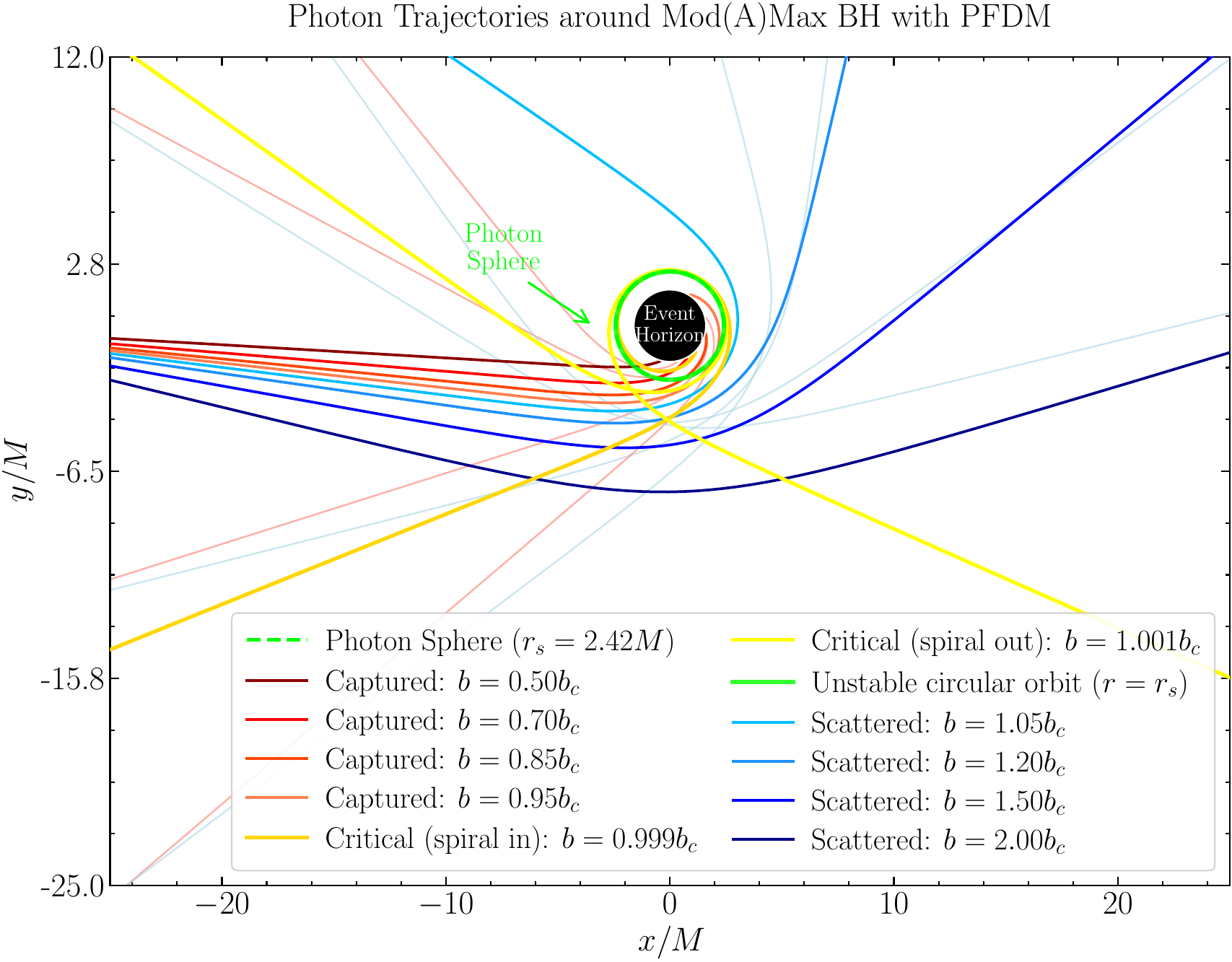}
    \caption{Complete photon trajectories around the Mod(A)Max black hole surrounded by perfect fluid dark matter, with parameters $Q = 0.5M$, $\gamma = 0.3$, $\lambda = 0.1$, and $\eta = +1$. The black filled circle represents the event horizon at $r_h = 1.61M$, and the green dashed circle marks the photon sphere at $r_s = 2.42M$. Red and orange curves show captured photons with impact parameters $b < b_c$ that spiral into the black hole. The yellow and gold curves represent near-critical orbits ($b \approx b_c = 4.17M$) that execute multiple revolutions around the photon sphere before either falling in or escaping. Blue curves depict scattered photons with $b > b_c$ that are deflected by the gravitational field but escape to infinity. The bright green curve on the photon sphere illustrates the unstable circular orbit where photons can theoretically orbit indefinitely. Light-colored trajectories show additional orbits from different incident angles, demonstrating the azimuthal symmetry of the scattering process.}
    \label{fig:photon_trajectories}
\end{figure}

By numerically integrating the orbit equation (\ref{bb16}), we can visualize the complete spectrum of photon trajectories around the black hole. The trajectories are classified into three distinct categories based on their impact parameter $b$ relative to the critical value $b_c$: captured orbits ($b < b_c$), critical orbits ($\beta \approx \beta_c$), and scattered orbits ($b> b_c$). Figure~\ref{fig:photon_trajectories} presents a comprehensive view of all three trajectory types.

To interpret Fig.~\ref{fig:photon_trajectories}, recall that the orbit equation (\ref{bb16}) can be written in the form of a ``radial motion'' with a turning point whenever the right-hand side vanishes. For $b<b_c$, no turning point occurs outside the horizon and trajectories cross the photon sphere and continue inward, producing the red/orange capture spirals. For $b>b_c$, the right-hand side becomes zero at a finite radius outside the photon sphere, generating a turning point: the photon approaches the black hole, reaches the periapsis, and then escapes to infinity, producing the blue scattering arcs.

The near-critical trajectories (yellow/gold) illustrate the key strong-field phenomenon: for $b \approx b_c$, the turning point occurs extremely close to the photon sphere and the photon executes many revolutions before deciding between capture or escape. This is the geometric origin of the photon ring: photons that linger near the unstable orbit accumulate large path length and strong focusing, contributing enhanced brightness in black hole images. The green circular orbit is shown as a reference: it is an unstable fixed point of the dynamical system, so any infinitesimal perturbation makes the photon spiral away (outwards) or spiral in (toward the horizon).

\begin{figure}[tbhp]
    \centering
    \includegraphics[width=\textwidth]{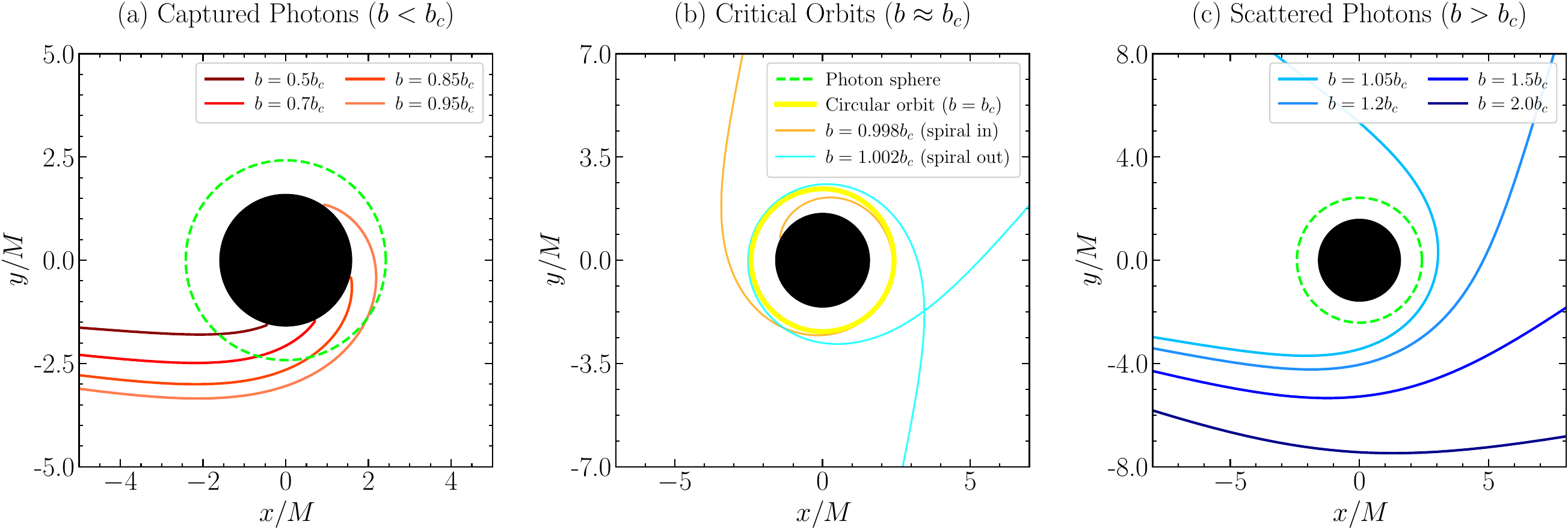}
    \caption{Classification of photon trajectories based on the impact parameter $b$ relative to the critical value $b_c$. Panel (a) shows captured photons ($b < b_c$): these trajectories cross the photon sphere and inevitably fall into the event horizon. The degree of deflection before capture increases as $\beta$ approaches $b_c$. Panel (b) illustrates critical and near-critical orbits ($b \approx b_c$): the yellow circle represents the theoretical unstable circular orbit at the photon sphere radius $r_s$. Orange and cyan curves show photons with impact parameters slightly below and above $b_c$, respectively, which spiral around the photon sphere multiple times before their ultimate fate is determined. These orbits are responsible for the photon ring observed in black hole images. Panel (c) displays scattered photons ($b > b_c$): these trajectories approach the black hole, reach a turning point outside the photon sphere, and escape to infinity. The deflection angle decreases as $b$ increases, asymptotically approaching zero for very large impact parameters.}
    \label{fig:three_cases}
\end{figure}

Figure~\ref{fig:three_cases} emphasizes the parametric control offered by the impact parameter. The three-panel visualization in Fig.~\ref{fig:three_cases} provides a clearer separation of the different trajectory types, allowing for detailed examination of each regime. In Fig. \ref{fig:three_cases}(a), moving from smaller to larger $b$ (still below $b_c$) increases the number of windings before capture because the photon approaches the unstable orbit more closely. Figure \ref{fig:three_cases}(b) demonstrates the separatrix-like behavior near $b_c$: tiny fractional changes in $\beta$ qualitatively switch the endpoint (capture versus escape), while producing a large number of quasi-circular loops. Figure \ref{fig:three_cases}(c) shows the weak-field limit emerging as $\beta$ increases: the periapsis moves outward, and the overall bending angle decreases, smoothly connecting to the Newtonian-like scattering regime.

\begin{figure}[ht!]
\centering
\includegraphics[scale=0.5]{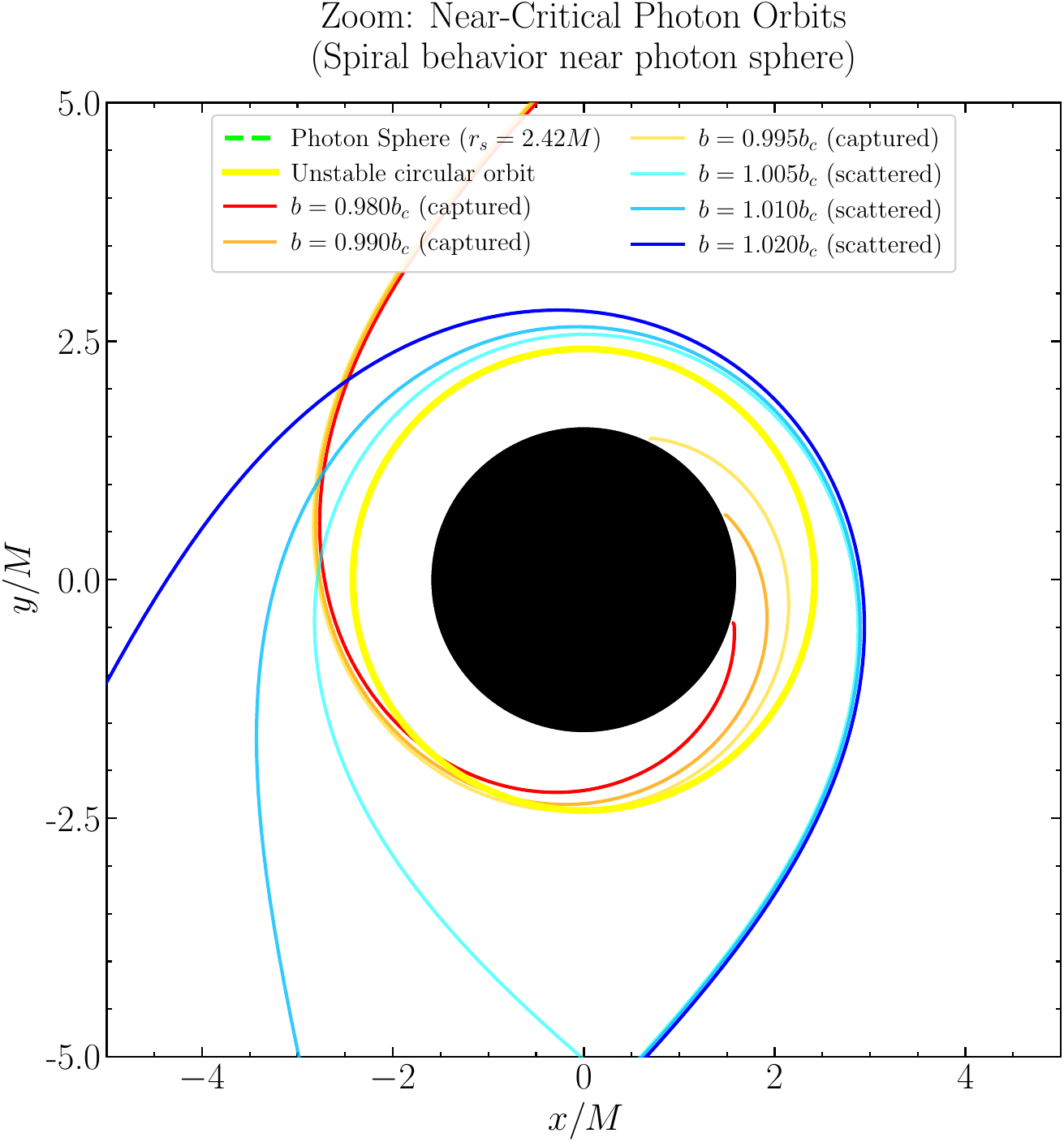}
\caption{Zoomed view of the photon sphere region showing the spiral behavior of near-critical photon orbits. The yellow circle represents the unstable circular orbit at $r = r_s = 2.42M$. Red, orange, and gold curves show photons with impact parameters slightly less than $b_c$ (specifically, $b = 0.98\,b_c$, $0.99\,b_c$, and $0.995\,b_c$), which spiral inward toward the event horizon after multiple orbits around the photon sphere. Cyan and blue curves represent photons with $b$ slightly greater than $b_c$ (specifically, $b = 1.005\,b_c$, $1.01\,b_c$, and $1.02\,b_c$), which spiral outward and eventually escape. The number of orbits around the photon sphere diverges logarithmically as $b \to b_c$, a characteristic feature of the unstable circular orbit. This behavior is responsible for the formation of higher-order photon rings in black hole images, with each successive ring corresponding to photons that complete an additional half-orbit around the black hole.}
    \label{fig:zoom_photon_sphere}
\end{figure}

The behavior of photon trajectories near the critical impact parameter is of particular interest, as it determines the characteristics of the photon ring-the bright ring of light observed in black hole images. Figure~\ref{fig:zoom_photon_sphere} provides a zoomed view of this region, highlighting the spiral dynamics that occur when $b \approx b_c$, which provides the most direct visualization of the instability of the circular null orbit. The key point is not only that trajectories ``wrap'' around $r=r_s$, but that the wrapping count increases sharply as $b$ approaches $b_c$ from either side. This implies that near-critical photons spend a long affine time near the photon sphere, amplifying observational signatures (multiple images, higher-order rings, and sharp brightness features). In computational terms, the spiraling also serves as a sensitive diagnostic: if the numerically integrated orbit exhibits many turns near $r_s$, then $b$ is very close to $\beta_c$ determined from Eq.~(\ref{bb15}).

\begin{figure}[ht!]
    \centering
    \includegraphics[width=\textwidth]{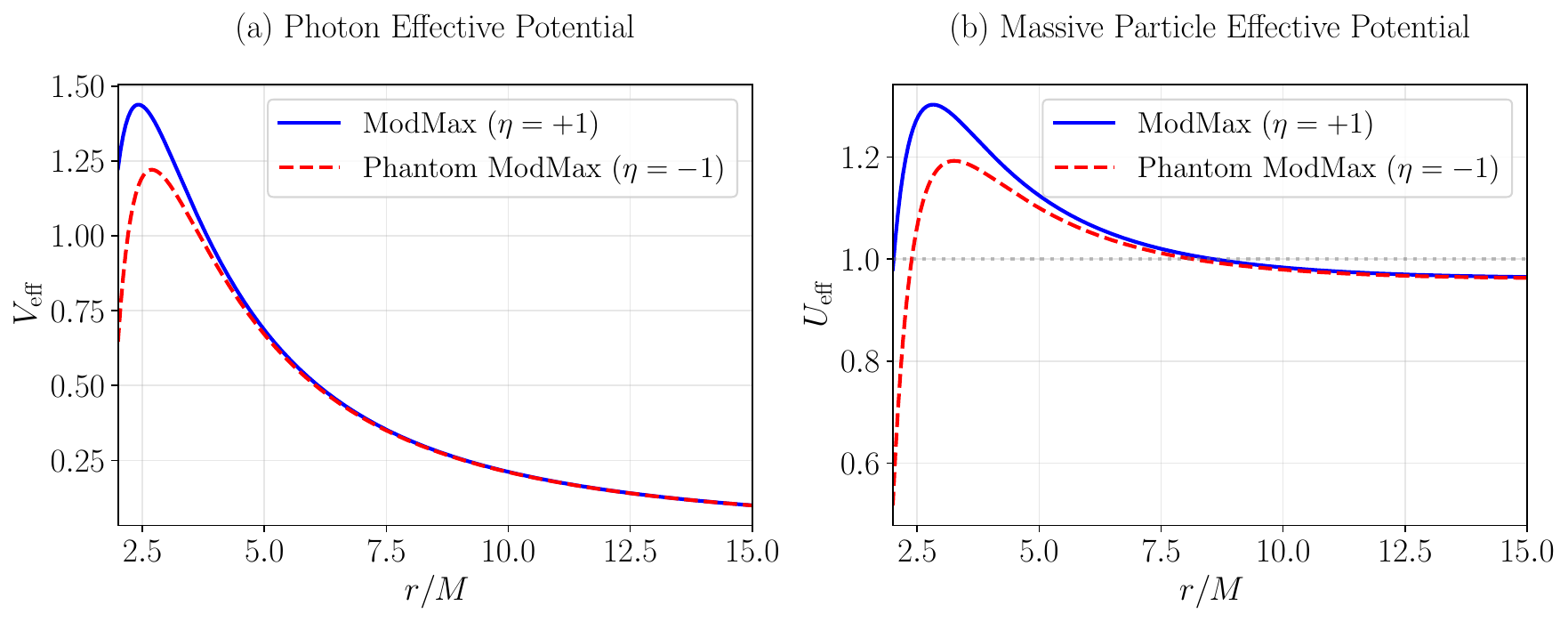}
    \caption{Comparison between ModMax ($\eta = +1$, blue curves) and phantom ModMax ($\eta = -1$, red dashed curves) black holes surrounded by perfect fluid dark matter. Panel (a) shows the photon effective potential $V_{\rm eff}$ for both configurations with $\mathrm{L} = 5M$, $Q = 0.5M$, $\gamma = 0.3$, and $\lambda = 0.1$. The ModMax case exhibits a higher potential barrier located at smaller radii compared to the phantom case, indicating a more compact photon sphere. Panel (b) displays the corresponding effective potential for massive particles with $\mathcal{L} = 4M$. The phantom configuration produces a shallower potential well with the minimum shifted to larger radii, implying that stable circular orbits for massive particles occur at greater distances from the event horizon. The horizontal dashed line at $U_{\rm eff} = 1$ represents the rest mass energy threshold, below which bound orbits are possible.}
    \label{fig:modmax_comparison}
\end{figure}

A comparison between the ModMax and phantom ModMax configurations is presented in Fig.~\ref{fig:modmax_comparison}, which illustrates how the sign of $\eta$ affects both the effective potential and the resulting photon dynamics.

Figure~\ref{fig:modmax_comparison} summarizes the qualitative impact of flipping $\eta$. In Fig. \ref{fig:modmax_comparison}(a), the phantom case lowers and shifts the photon barrier outward, consistent with the idea that the sign reversal reduces the ``Reissner-Nordstr\"om-like'' repulsive contribution in $f(r)$. Observationally, this tends to enlarge the characteristic optical scales (photon sphere radius and shadow size) relative to the $\eta=+1$ case for the same $(Q,\gamma,\lambda)$. Figure \ref{fig:modmax_comparison}(b) shows that the same sign flip also weakens the depth of the massive-particle potential well and moves its minimum outward, which implies that characteristic orbital radii (including the ISCO) are expected to shift to larger $r$ in the phantom configuration. This is relevant for accretion modeling: shifting stable circular orbits outward typically reduces binding energy release and can impact radiative efficiency estimates.

Finally, we determine the effective radial force experiences by the photons in the given gravitational field produced by Mod(A)Max black hole with PFDM. This force demonstrate whether the photons is captured by the black hole strong field or escape away to infinity. In terms of effective potential of null geodesics, the effective radial force is the negative gradient of the potential. Mathematically, it is defined by
\begin{equation}
    \mathcal{F}_{\rm rad}=-\frac{1}{2}\frac{\partial V_{\rm eff}}{\partial r}.\label{force1}
\end{equation}
Substituting the potential $V_{\rm eff}$ given in (\ref{bb7}) and after simplification yields
\begin{equation}
\mathcal{F}_{\rm rad}=\frac{\mathrm{L}^2}{r^3}\left(1-\frac{3 M}{r}+2\eta\,e^{-\gamma}\,\frac{Q^2}{r^2}-\frac{\lambda}{2r}+\frac{3\lambda}{2r}\ln\!\frac{r}{|\lambda|}\right).\label{force2}
\end{equation}

\begin{figure}[t]
\centering
\includegraphics[width=0.45\linewidth]{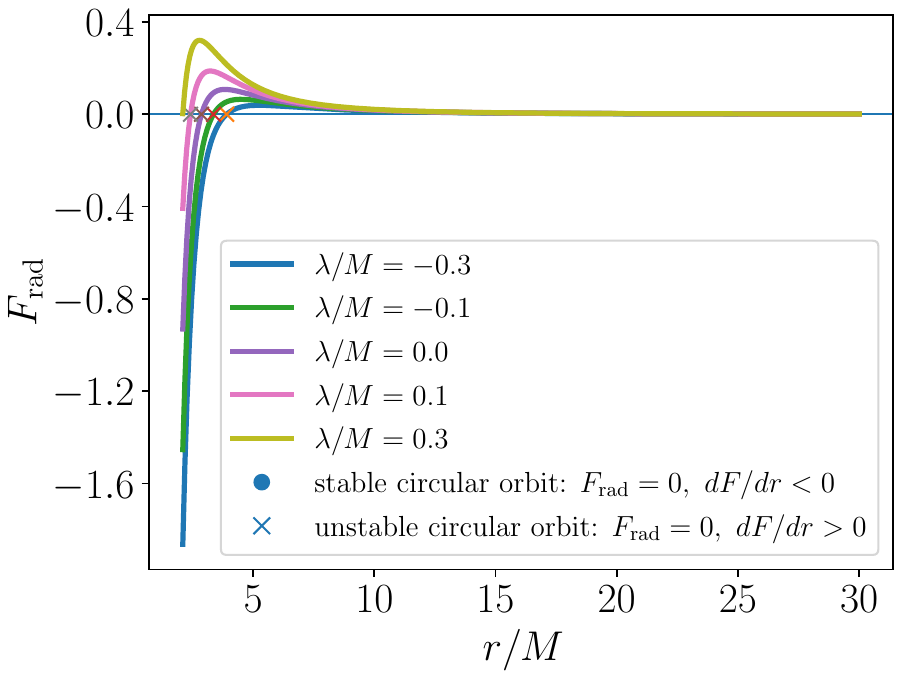}
\caption{Effective radial force for photons, $F_{\rm rad}(r)$, obtained from Eq.~(28) for different values of the PFDM parameter $\lambda/M=\{-0.3,-0.1,0,0.1,0.3\}$, with fixed $M$, $Q$, $\gamma$, and $\eta$ (see text). The circular photon orbits are identified by the zeros of $F_{\rm rad}(r)$ (marked points). Stable circular orbits are indicated by open circles ($F_{\rm rad}=0$ and $dF_{\rm rad}/dr<0$), whereas unstable circular orbits are indicated by crosses ($F_{\rm rad}=0$ and $dF_{\rm rad}/dr>0$).}
    \label{fig:Frad_photons_eq28}
\end{figure}
Figure~\ref{fig:Frad_photons_eq28} shows the photon effective radial force associated with Eq.~(28). 
For each $\lambda$, the circular photon orbits are determined by the roots of $F_{\rm rad}(r)$, i.e., the intersections with the horizontal axis. 
The stability of each circular orbit is inferred from the local slope: a negative slope ($dF_{\rm rad}/dr<0$) yields a restoring behavior against small radial perturbations, characterizing a stable circular orbit, whereas a positive slope ($dF_{\rm rad}/dr>0$) characterizes an unstable orbit. 
The PFDM parameter $\lambda$ shifts the location (and potentially the multiplicity) of these roots, thereby deforming the photon orbital structure when compared with the $\lambda=0$ case.

From the expression (\ref{force2}), it is evident that the black hole mass $M$, the electric charge $Q$, the ModMax parameter $\gamma$, and the PFDM parameter $\lambda$ significantly affect the effective radial force experienced by photons. The conserved angular momentum $\mathrm{L}$ of the photons also plays a crucial role. Since the radial force $\mathcal{F}_{\rm rad}$ vanishes at the photon-sphere radius $r_s$, plotting $\mathcal{F}_{\rm rad}$ as a function of the radial coordinate $r$ for suitable choices of the parameters allows one to clearly identify the regions corresponding to photons pull and push.   
\begin{table}[ht!]
\centering
\footnotesize
\caption{Event horizon $r_h$, photon sphere $r_s$, and shadow radius $R_{sh}$ for different $\gamma$ and $\lambda$ values ($Q=0.1$).}
\begin{tabular}{|c|c|c|c|c||c|c|c|}
\hline
\multicolumn{5}{|c||}{$\eta=+1$} & \multicolumn{3}{c|}{$\eta=-1$} \\
\hline
$\gamma$ & $\lambda/M$ & $r_h/M$ & $r_s/M$ & $R_{sh}/M$ & $r_h/M$ & $r_s/M$ & $R_{sh}/M$ \\
\hline
0.1 & 0.1 & 1.71076 & 2.55673 & 4.34713 & 1.72073 & 2.57006 & 4.36445 \\
0.1 & 0.3 & 1.50932 & 2.23768 & 3.64238 & 1.51929 & 2.25111 & 3.65973 \\
0.1 & 0.5 & 1.45851 & 2.14823 & 3.35390 & 1.46773 & 2.16069 & 3.36986 \\
0.1 & 0.7 & 1.47305 & 2.15899 & 3.25198 & 1.48136 & 2.17025 & 3.26625 \\
0.1 & 0.9 & 1.52150 & 2.22186 & 3.24871 & 1.52896 & 2.23197 & 3.26140 \\
0.3 & 0.1 & 1.71167 & 2.55794 & 4.34871 & 1.71983 & 2.56886 & 4.36289 \\
0.3 & 0.3 & 1.51023 & 2.23890 & 3.64396 & 1.51839 & 2.24990 & 3.65817 \\
0.3 & 0.5 & 1.45935 & 2.14937 & 3.35535 & 1.46690 & 2.15957 & 3.36842 \\
0.3 & 0.7 & 1.47381 & 2.16001 & 3.25328 & 1.48061 & 2.16923 & 3.26496 \\
0.3 & 0.9 & 1.52218 & 2.22278 & 3.24987 & 1.52829 & 2.23106 & 3.26026 \\
0.5 & 0.1 & 1.71241 & 2.55893 & 4.35000 & 1.71909 & 2.56787 & 4.36161 \\
0.5 & 0.3 & 1.51097 & 2.23990 & 3.64525 & 1.51765 & 2.24891 & 3.65688 \\
0.5 & 0.5 & 1.46004 & 2.15029 & 3.35654 & 1.46622 & 2.15865 & 3.36724 \\
0.5 & 0.7 & 1.47443 & 2.16085 & 3.25434 & 1.48000 & 2.16840 & 3.26391 \\
0.5 & 0.9 & 1.52274 & 2.22353 & 3.25081 & 1.52774 & 2.23031 & 3.25932 \\
0.7 & 0.1 & 1.71302 & 2.55975 & 4.35105 & 1.71849 & 2.56707 & 4.36056 \\
0.7 & 0.3 & 1.51158 & 2.24072 & 3.64631 & 1.51705 & 2.24809 & 3.65583 \\
0.7 & 0.5 & 1.46060 & 2.15105 & 3.35751 & 1.46566 & 2.15789 & 3.36627 \\
0.7 & 0.7 & 1.47493 & 2.16154 & 3.25521 & 1.47949 & 2.16772 & 3.26304 \\
0.7 & 0.9 & 1.52319 & 2.22414 & 3.25158 & 1.52728 & 2.22970 & 3.25855 \\
0.9 & 0.1 & 1.71351 & 2.56041 & 4.35192 & 1.71799 & 2.56640 & 4.35970 \\
0.9 & 0.3 & 1.51207 & 2.24139 & 3.64717 & 1.51656 & 2.24743 & 3.65497 \\
0.9 & 0.5 & 1.46106 & 2.15168 & 3.35831 & 1.46520 & 2.15727 & 3.36548 \\
0.9 & 0.7 & 1.47535 & 2.16210 & 3.25592 & 1.47908 & 2.16716 & 3.26233 \\
0.9 & 0.9 & 1.52356 & 2.22465 & 3.25222 & 1.52691 & 2.22920 & 3.25792 \\
\hline
\end{tabular}
\label{tab:radii1}
\end{table}

\begin{table}[ht!]
\centering
\footnotesize
\caption{Event horizon $r_h$, photon sphere $r_s$, and shadow radius $R_{sh}$ for different $\gamma$ and $\lambda$ values ($Q=0.3$).}
\begin{tabular}{|c|c|c|c|c||c|c|c|}
\hline
\multicolumn{5}{|c||}{$\eta=+1$} & \multicolumn{3}{c|}{$\eta=-1$} \\
\hline
$\gamma$ & $\lambda/M$ & $r_h/M$ & $r_s/M$ & $R_{sh}/M$ & $r_h/M$ & $r_s/M$ & $R_{sh}/M$ \\
\hline
0.1 & 0.1 & 1.66970 & 2.50195 & 4.27625 &  1.75952 & 2.62213 & 4.43225 \\
0.1 & 0.3 & 1.46814 & 2.18239 & 3.57120 &  1.55805 & 2.30344 & 3.72755 \\
0.1 & 0.5 & 1.42057 & 2.09707 & 3.28859 &  1.50364 & 2.20934 & 3.43232 \\
0.1 & 0.7 & 1.43898 & 2.11293 & 3.19378 &  1.51386 & 2.21434 & 3.32226 \\
0.1 & 0.9 & 1.49104 & 2.18061 & 3.19711 &  1.55824 & 2.27173 & 3.31136 \\
0.3 & 0.1 & 1.67824 & 2.51332 & 4.29092 &  1.75174 & 2.61167 & 4.41861 \\
0.3 & 0.3 & 1.47671 & 2.19388 & 3.58595 &  1.55028 & 2.29294 & 3.71392 \\
0.3 & 0.5 & 1.42845 & 2.10768 & 3.30211 &  1.49644 & 2.19957 & 3.41975 \\
0.3 & 0.7 & 1.44604 & 2.12246 & 3.20580 &  1.50733 & 2.20547 & 3.31097 \\
0.3 & 0.9 & 1.49733 & 2.18912 & 3.20775 &  1.55234 & 2.26371 & 3.30127 \\
0.5 & 0.1 & 1.68516 & 2.52254 & 4.30285 &  1.74532 & 2.60305 & 4.40737 \\
0.5 & 0.3 & 1.48366 & 2.20320 & 3.59794 &  1.54388 & 2.28428 & 3.70268 \\
0.5 & 0.5 & 1.43485 & 2.11630 & 3.31310 &  1.49050 & 2.19151 & 3.40940 \\
0.5 & 0.7 & 1.45178 & 2.13021 & 3.21558 &  1.50194 & 2.19816 & 3.30168 \\
0.5 & 0.9 & 1.50245 & 2.19605 & 3.21641 &  1.54748 & 2.25711 & 3.29298 \\
0.7 & 0.1 & 1.69079 & 2.53005 & 4.31256 &  1.74004 & 2.59595 & 4.39812 \\
0.7 & 0.3 & 1.48931 & 2.21078 & 3.60769 &  1.53859 & 2.27715 & 3.69344 \\
0.7 & 0.5 & 1.44005 & 2.12331 & 3.32204 &  1.48560 & 2.18488 & 3.40088 \\
0.7 & 0.7 & 1.45644 & 2.13652 & 3.22355 &  1.49751 & 2.19214 & 3.29404 \\
0.7 & 0.9 & 1.50663 & 2.20170 & 3.22348 &  1.54349 & 2.25169 & 3.28616 \\
0.9 & 0.1 & 1.69537 & 2.53616 & 4.32047 &  1.73568 & 2.59011 & 4.39052 \\
0.9 & 0.3 & 1.49390 & 2.21694 & 3.61563 &  1.53425 & 2.27128 & 3.68583 \\
0.9 & 0.5 & 1.44428 & 2.12902 & 3.32933 &  1.48157 & 2.17942 & 3.39388 \\
0.9 & 0.7 & 1.46025 & 2.14166 & 3.23005 &  1.49387 & 2.18720 & 3.28776 \\
0.9 & 0.9 & 1.51003 & 2.20631 & 3.22924 &  1.54021 & 2.24723 & 3.28056 \\
\hline
\end{tabular}
\label{tab:radii2}
\end{table}

\section{Dynamics of Massive Particles}

In this section, we study the dynamics of massive test particles around the Mod(A)Max black hole surrounded by a perfect fluid dark matter. We explore the specific angular momentum and specific energy of test particles revolving around the black hole in circular orbits of fixed radius showing the effects of geometric parameters on them. Moreover, using stability conditions for circular orbits, we determine innermost stable circular orbits (ISCO) and analyze how the various parameters influence on this. 

The Lagrangian density function for test particle of mass $m$ is given by
\begin{equation}
\mathbb{L}=\frac{1}{2}\,m\,g_{\mu\nu}\, \dot x^{\mu}\,\dot x^{\nu},\label{cc1}
\end{equation}
Following the previous procedure, one can find the following set of equations of motion for massive test particles on the equatorial plane as,
\begin{align}
&p_t/m=-\left(1-\frac{2 M}{r}+\eta\,\frac{e^{-\gamma}\,Q^2}{r^2}+\frac{\lambda}{r}\ln\!\frac{r}{|\lambda|}\right)\,\frac{dt}{d\zeta}=-\mathcal{E},\label{cc2}\\
&p_{\phi}/m=r^2 \frac{d\phi}{d\zeta}=\mathcal{L},\label{cc3}\\
&\left(\frac{dr}{d\zeta}\right)^2+\left(1+\frac{\mathcal{L}^2}{r^2}\right)\,\left(1-\frac{2 M}{r}+\eta\,\frac{e^{-\gamma}\,Q^2}{r^2}+\frac{\lambda}{r}\ln\!\frac{r}{|\lambda|}\right)=\mathcal{E}^2.\label{cc4}
\end{align}

Therefore, the first integral equation of motion of massive particles on the equatorial plane is given by
\begin{align}
&\frac{dt}{d\zeta}=\mathcal{E}\,\left(1-\frac{2 M}{r}+\eta\,\frac{e^{-\gamma}\,Q^2}{r^2}+\frac{\lambda}{r}\ln\!\frac{r}{|\lambda|}\right)^{-1},\label{cc2a}\\
&\frac{d\phi}{d\zeta}=\mathcal{L}/r^2,\label{cc3a}\\
&\frac{dr}{d\zeta}=\sqrt{\mathcal{E}^2-\left(1+\frac{\mathcal{L}^2}{r^2}\right)\,\left(1-\frac{2 M}{r}+\eta\,\frac{e^{-\gamma}\,Q^2}{r^2}+\frac{\lambda}{r}\ln\!\frac{r}{|\lambda|}\right)}.\label{cc4a}
\end{align}

Equation (\ref{cc4a}) is equivalent to the one-dimensional equation of motion of test particle in the form $\left(\frac{dr}{d\zeta}\right)^2+U_{\rm eff}(r)=\mathcal{E}^2$, where $U_{\rm eff}(r)$ is the effective potential that governs the particle dynamics and is given by
\begin{equation}
U_{\rm eff}(r)=\left(1+\frac{\mathcal{L}^2}{r^2}\right)\,\left(1-\frac{2 M}{r}+\eta\,\frac{e^{-\gamma}\,Q^2}{r^2}+\frac{\lambda}{r}\ln\!\frac{r}{|\lambda|}\right).\label{cc5}
\end{equation}

\begin{figure}[ht!]
    \centering
    \includegraphics[width=\textwidth]{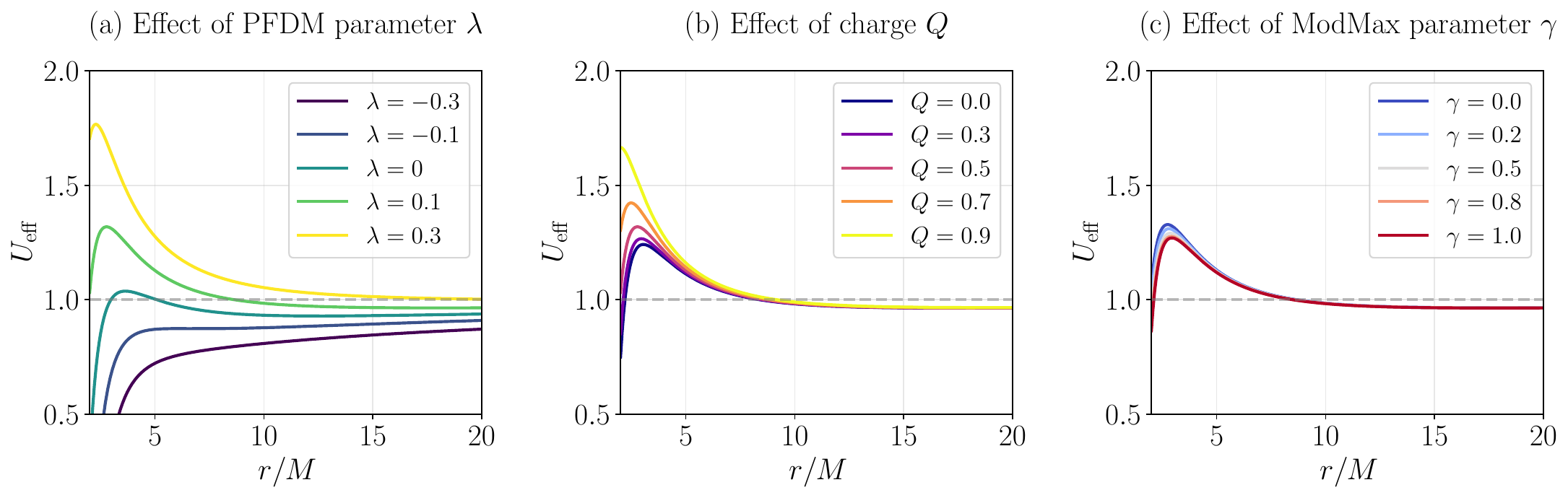}
    \caption{The effective potential $U_{\rm eff}$ for massive particle motion as a function of the radial coordinate $r/M$ in the Mod(A)Max black hole spacetime surrounded by perfect fluid dark matter. The specific angular momentum is set to $\mathcal{L} = 4M$. Panel (a) shows the influence of the PFDM parameter $\lambda$: negative values deepen the potential well and shift the minimum to larger radii, while positive values have the opposite effect. Panel (b) illustrates the dependence on the electric charge $Q$; increasing the charge raises the potential barrier near the horizon and shifts the minimum outward, providing stronger binding for orbiting particles. Panel (c) demonstrates the effect of the ModMax parameter $\gamma$, where larger values suppress the electromagnetic contribution and lead to a shallower potential structure. The horizontal dashed line at $U_{\rm eff} = 1$ indicates the threshold for bound orbits: particles with $\mathcal{E} < 1$ are gravitationally bound, while those with $\mathcal{E} > 1$ can escape to infinity.}
    \label{fig:massive_potential}
\end{figure}

The effective potential for massive particles differs fundamentally from the photon case due to the additional rest mass contribution. Figure~\ref{fig:massive_potential} presents a detailed analysis of how this potential varies with the model parameters.

Figure~\ref{fig:massive_potential} encodes the orbital taxonomy for massive particles. For a given $\mathcal{L}$, a local \emph{minimum} of $U_{\rm eff}$ corresponds to a stable circular orbit, while a local \emph{maximum} corresponds to an unstable circular orbit that separates bound motion from plunging motion. The reference line $U_{\rm eff}=1$ marks the rest-mass energy of a particle at infinity: if $\mathcal{E}<1$, the particle is bound and oscillates between turning points; if $\mathcal{E}>1$, it may have unbound trajectories.  Figure \ref{fig:massive_potential}(a) shows that PFDM can deepen the potential well (negative $\lambda$) and move the stable orbit region outward, altering both the location and depth of the minimum. This has direct implications for accretion disk modeling, because the characteristic radii of circular motion and the energy release (binding energy) depend on where the minimum occurs. Figure \ref{fig:massive_potential}(b) indicates that increasing $Q$ modifies the near-horizon structure and reshapes the potential; for the displayed range, the minimum moves outward, which signals that stable circular orbits are displaced to larger radii. Figure \ref{fig:massive_potential}(c) shows that increasing $\gamma$ suppresses electromagnetic effects and makes the potential increasingly dominated by the PFDM-deformed Schwarzschild-like term, again explaining why strong charge signatures can be reduced by large $\gamma$.

Next, we determine the trajectories of the massive test particles in the given gravitational field and analyze how geometric parameters influence the particles trajectory. Using Eqs.~(\ref{cc3}) and (\ref{cc4}), we obtain
\begin{equation}
\left(\frac{dr}{d\phi}\right)^2=\frac{\mathcal{E}^2}{\mathcal{L}^2}-\left(\frac{1}{\mathcal{L}^2}+\frac{1}{r^2}\right)\,\left(1-\frac{2 M}{r}+\eta\,\frac{e^{-\gamma}\,Q^2}{r^2}+\frac{\lambda}{r}\ln\!\frac{r}{|\lambda|}\right).\label{trajectory}
\end{equation}
From the above expression (\ref{trajectory}), it is evident that the black hole mass $M$, the electric charge $Q$, the ModMax parameter $\gamma$, and the PFDM parameter $\lambda$ significantly affect the particles trajectory traversing in the field. 

\begin{figure}[ht!]
\centering
\includegraphics[width=0.95\linewidth]{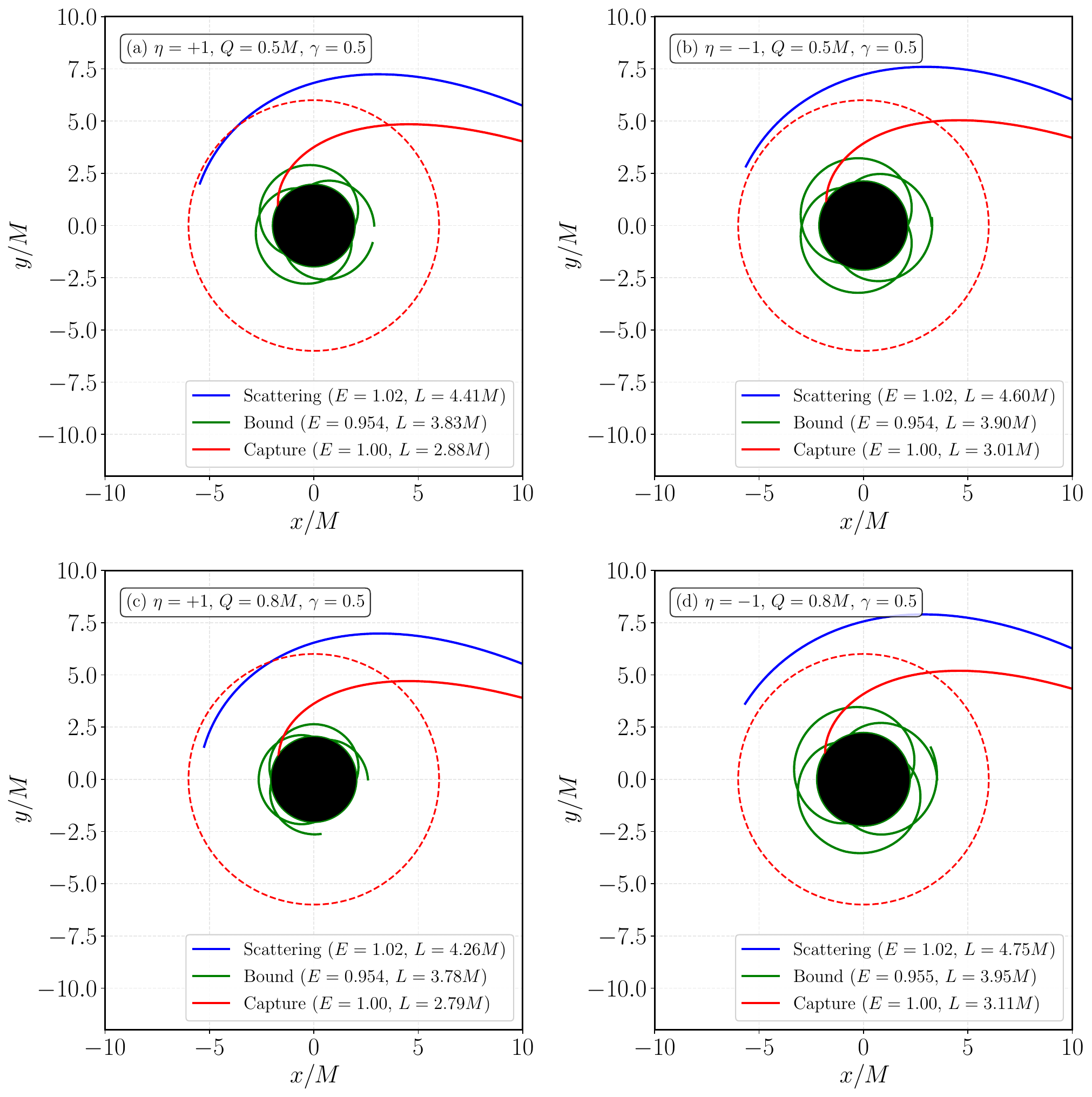}
\caption{Timelike geodesics of massive particles around the Mod(A)Max black hole for three different regimes: scattering (blue), bound (green), and capture (red). The black filled circle represents the event horizon $r_h$, and the red dashed circle indicates the innermost stable circular orbit (ISCO). Panels (a) and (b) correspond to $Q = 0.5M$, while panels (c) and (d) show $Q = 0.8M$. The left column displays the ModMax case ($\eta = +1$) and the right column the Mod(A)Max case ($\eta = -1$). In all panels, $M = 1$ and $\gamma = 0.5$. The bound orbits exhibit perihelion precession characteristic of non-Keplerian potentials.}
\label{fig:massive_trajectory}
\end{figure}
In Fig.~\ref{fig:massive_trajectory}, we display the trajectories of massive test particles (Eq. (\ref{trajectory})) around the Mod(A)Max black hole for three distinct dynamical regimes. The scattering regime (blue curves) corresponds to particles with specific energy $\mathcal{E} > 1$, which approach the black hole from infinity, experience gravitational deflection, and escape back to infinity along hyperbolic-like trajectories. The bound regime (green curves) represents particles with $\mathcal{E} < 1$ and angular momentum $\mathcal{L} > \mathcal{L}_{\rm ISCO}$, which remain gravitationally bound to the black hole in stable elliptical-like orbits. The capture regime (red curves) corresponds to particles with angular momentum below the critical value $\mathcal{L} < \mathcal{L}_c$, which spiral inward and inevitably cross the event horizon.

A distinctive feature of the bound orbits is the perihelion precession, clearly visible in the green trajectories. Unlike Newtonian gravity, where bound orbits close after one period, the relativistic corrections in the Mod(A)Max spacetime cause the perihelion to advance with each orbit. This precession is a well-known general relativistic effect, famously confirmed by Mercury's orbital precession, and its magnitude depends on the spacetime parameters $Q$, $\gamma$, and $\eta$.

Comparing the left and right columns, we observe that the event horizon radius is systematically larger in the Mod(A)Max case ($\eta = -1$) than in the ModMax case ($\eta = +1$) for the same charge $Q$. This is consistent with our previous analysis showing that the phantom-like contribution ($\eta = -1$) effectively increases the gravitational attraction. Consequently, the capture cross-section is enhanced in the Mod(A)Max scenario, as evidenced by the red trajectories reaching the horizon at larger radial distances.

The ISCO, indicated by the red dashed circle, marks the boundary between stable and unstable circular orbits. Particles with $\mathcal{L} < \mathcal{L}_{\rm ISCO}$ cannot maintain stable circular motion and will either plunge into the black hole or escape to infinity, depending on their energy. The position of the ISCO is particularly relevant for accretion disk physics, as it determines the inner edge of geometrically thin disks and influences the spectral properties of the emitted radiation.

These results demonstrate that the interplay between the electric charge $Q$ and the nonlinear electrodynamic parameter $\eta$ significantly affects the dynamics of massive particles around the black hole. The distinct orbital characteristics in the ModMax and Mod(A)Max scenarios could, in principle, lead to observable differences in the properties of accretion disks and the orbital motion of stars in the vicinity of charged compact objects.

In addition, we calculate the effective radial force experiences by the massive test particles and analyze the effects of geometric parameters that changes the space-time curvature. Mathematically, this effective radial force  is defined by
\begin{equation}
    \mathbb{F}_{\rm rad}=-\frac{1}{2}\frac{\partial U_{\rm eff}}{\partial r}=-\left(\frac{M}{r^2}-\eta\,e^{-\gamma}\,\frac{Q^2}{r^3}+\frac{\lambda}{2r^2}-\frac{\lambda}{2r^2}\ln\!\frac{r}{|\lambda|}\right)+\frac{\mathcal{L}^2}{r^3}\,\left(1-\frac{3 M}{r}+2\eta\,e^{-\gamma}\,\frac{Q^2}{r^2}-\frac{\lambda}{2r}+\frac{3\lambda}{2r}\ln\!\frac{r}{|\lambda|}\right).\label{particle-force}
\end{equation}
\begin{figure}[tbhp]
\centering
\includegraphics[width=0.45\linewidth]{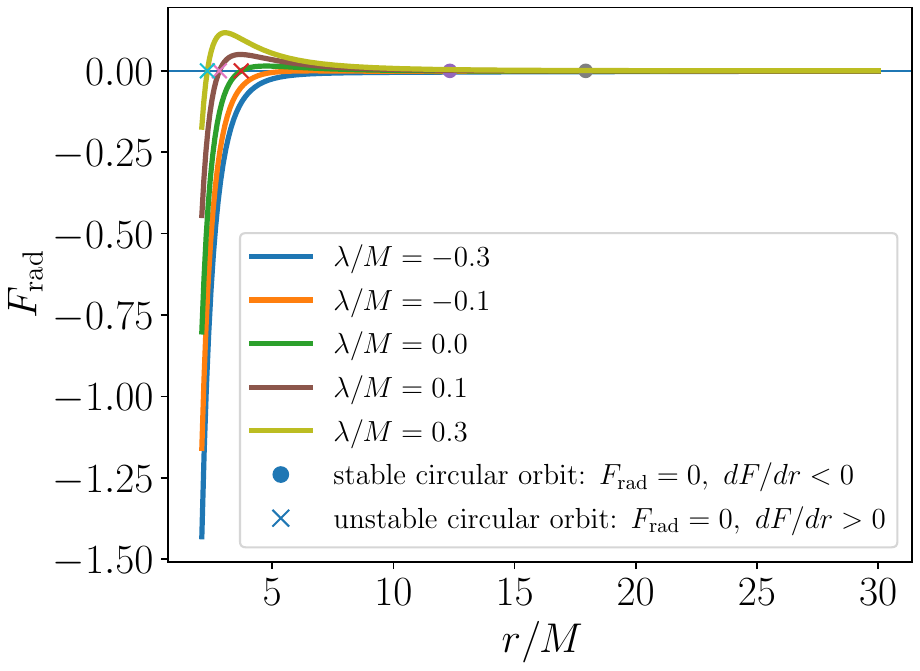}
    \caption{Effective radial force experienced by massive particles, $\mathbb{F}_{\rm rad}(r)$, obtained from Eq.~(\ref{particle-force}) for $\lambda/M=\{-0.3,-0.1,0,0.1,0.3\}$, with the remaining parameters fixed as in Fig.~\ref{fig:Frad_photons_eq28} (see text). Circular massive-particle orbits correspond to the zeros of $\mathbb{F}_{\rm rad}(r)$ (marked points). Stable circular orbits are denoted by open circles ($\mathbb{F}_{\rm rad}=0$ and $d\mathbb{F}_{\rm rad}/dr<0$), while unstable circular orbits are denoted by crosses ($\mathbb{F}_{\rm rad}=0$ and $d\mathbb{F}_{\rm rad}/dr>0$).}
    \label{fig:Frad_massive_eq38}
\end{figure}
In Fig.~\ref{fig:Frad_massive_eq38} we depict the massive-particle effective radial force dictated by Eq.~(\ref{particle-force}). 
As in the photon case, circular orbits occur at the radii satisfying $\mathbb{F}_{\rm rad}(r)=0$, and their stability follows from the sign of the derivative at the root: $d\mathbb{F}_{\rm rad}/dr<0$ indicates stability, whereas $d\mathbb{F}_{\rm rad}/dr>0$ indicates instability. 
Because Eq.~(\ref{particle-force}) contains, in addition to the $\mathcal{L}^2/r^3$ contribution, a term that resembles a Newtonian-like attraction/repulsion, the overall balance that produces the roots is more sensitive to parameter changes. 
Consequently, varying $\lambda$ not only shifts the radii of circular orbits but can also modify the radial intervals where stable motion is allowed, which is relevant for identifying transitions in the orbital structure (e.g., the onset/disappearance of stable branches).

From the expression (\ref{particle-force}), it is evident that the black hole mass $M$, the electric charge $Q$, the ModMax parameter $\gamma$, and the PFDM parameter $\lambda$ significantly affect the effective radial force experienced by massive test particles traversing around the black hole. Moreover, the conserved angular momentum $\mathcal{L}$ of particles also plays a crucial role.

For circular orbits, the following conditions must be satisfied:
\begin{equation}
    U_{\rm eff}(r)=\mathcal{E}^2,\qquad \partial_r U_{\rm eff}(r)=0.\label{cc6}
\end{equation}
Substituting the effective potential in Eq. (\ref{cc5}) into the Eq. (\ref{cc6}) and after simplification one can find the specific angular momentum ($\mathcal{L}_{\rm sp}$) and specific energy ($\mathcal{E}_{\rm sp}$) of massive test particles orbiting around black holes in fixed orbits. These are given by
\begin{equation}
    \mathcal{L}_{\rm sp}=r\,\sqrt{\frac{\frac{M}{r}
- \frac{\eta\, e^{-\gamma} Q^2}{r^2}
+ \frac{\lambda}{2r}\left(1-\ln\! \frac{r}{|\lambda|}\right)}{1-\frac{3M}{r}
+ \frac{2\eta\, e^{-\gamma}Q^2}{r^2}
+ \frac{\lambda}{2r}\left(3\ln\! \frac{r}{|\lambda|}-1\right)}}.\label{cc7}
\end{equation}
And
\begin{equation}
    \mathcal{E}_{\rm sp}=\pm\,\frac{\left(1-\frac{2 M}{r}+\eta\,\frac{e^{-\gamma}\,Q^2}{r^2}+\frac{\lambda}{r}\ln\!\frac{r}{|\lambda|}\right)}{\sqrt{1-\frac{3M}{r}
+ \frac{2\eta\, e^{-\gamma}Q^2}{r^2}
+ \frac{\lambda}{2r}\left(3\ln\! \frac{r}{|\lambda|}-1\right)}}.\label{cc8}
\end{equation}

From the expressions (\ref{cc7})--(\ref{cc8}), it becomes evident that the specific angular momentum and specific energy are influenced by the electric charge $Q$ of the black hole, the PFDM parameter $\lambda$, and the ModMax parameter \(\gamma\). 
\begin{itemize}
    \item For $\eta=+1$, corresponding to the ModMax black hole with PFDM, these physical quantities simplify as,
    \begin{equation}
        \mathcal{L}_{\rm sp}=r\,\sqrt{\frac{\frac{M}{r}
- \frac{e^{-\gamma} Q^2}{r^2}
+ \frac{\lambda}{2r}\left(1-\ln\! \frac{r}{|\lambda|}\right)}{1-\frac{3M}{r}
+ \frac{2e^{-\gamma}Q^2}{r^2}
+ \frac{\lambda}{2r}\left(3\ln\! \frac{r}{|\lambda|}-1\right)}},\quad \mathcal{E}_{\rm sp}=\pm\,\frac{\left(1-\frac{2 M}{r}+\frac{e^{-\gamma}\,Q^2}{r^2}+\frac{\lambda}{r}\ln\!\frac{r}{|\lambda|}\right)}{\sqrt{1-\frac{3M}{r}
+ \frac{2e^{-\gamma}Q^2}{r^2}
+ \frac{\lambda}{2r}\left(3\ln\! \frac{r}{|\lambda|}-1\right)}}.\label{cc8a}
    \end{equation}
    \item For $\eta=-1$ corresponding to the phantom ModMax or Mod(A)Max black hole with PFDM, these physical quantities simplify as,
   \begin{equation}
       \mathcal{L}_{\rm sp}=r\,\sqrt{\frac{\frac{M}{r}
+\frac{e^{-\gamma} Q^2}{r^2}
+ \frac{\lambda}{2r}\left(1-\ln\! \frac{r}{|\lambda|}\right)}{1-\frac{3M}{r}
- \frac{2e^{-\gamma}Q^2}{r^2}
+ \frac{\lambda}{2r}\left(3\ln\! \frac{r}{|\lambda|}-1\right)}},\quad \mathcal{E}_{\rm sp}=\pm\,\frac{\left(1-\frac{2 M}{r}-\frac{e^{-\gamma}\,Q^2}{r^2}+\frac{\lambda}{r}\ln\!\frac{r}{|\lambda|}\right)}{\sqrt{1-\frac{3M}{r}
- \frac{2e^{-\gamma}Q^2}{r^2}
+ \frac{\lambda}{2r}\left(3\ln\! \frac{r}{|\lambda|}-1\right)}}.\label{cc8b}
   \end{equation}
\end{itemize}

\begin{figure}[ht!]
    \centering
    \includegraphics[width=\textwidth]{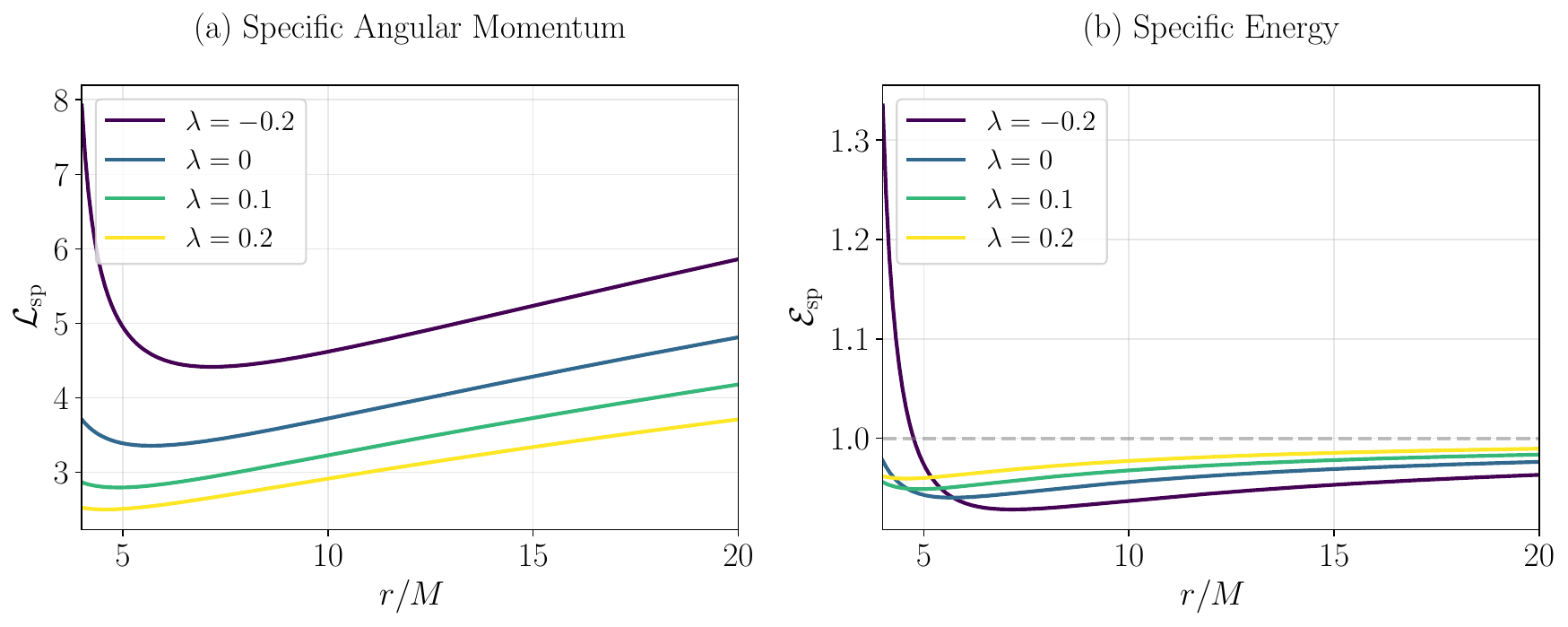}
    \caption{The specific angular momentum $\mathcal{L}_{\rm sp}$ and specific energy $\mathcal{E}_{\rm sp}$ for circular orbits around the Mod(A)Max black hole surrounded by perfect fluid dark matter. Panel (a) displays $\mathcal{L}_{\rm sp}$ as a function of radial distance for different values of the PFDM parameter $\lambda$. Negative values of $\lambda$ increase the angular momentum required for circular orbits at a given radius, while positive values decrease it. The curves diverge at the innermost stable circular orbit (ISCO), indicating the minimum radius for stable circular motion. Panel (b) shows the corresponding specific energy $\mathcal{E}_{\rm sp}$, which approaches unity at large radii (representing particles at rest at infinity). The energy decreases as particles orbit closer to the black hole, reaching a minimum at the ISCO before becoming undefined at smaller radii where no stable circular orbits exist. The horizontal dashed line at $\mathcal{E}_{\rm sp} = 1$ marks the binding threshold; orbits with $\mathcal{E}_{\rm sp} < 1$ are gravitationally bound, with the binding energy given by $1 - \mathcal{E}_{\rm sp}$.}
    \label{fig:specific_quantities}
\end{figure}

The radial profiles of the specific angular momentum and specific energy are shown in Fig.~\ref{fig:specific_quantities}. These quantities are essential for understanding the orbital characteristics of accretion disks around black holes.

Figure~\ref{fig:specific_quantities} provides a direct diagnostic of disk-relevant radii. The divergence of $\mathcal{L}_{\rm sp}$ in Fig. \ref{fig:specific_quantities}(a) indicates the breakdown of stable circular orbits: as one approaches the ISCO from above, maintaining a circular orbit would require arbitrarily large angular momentum, so physical (finite-$\mathcal{L}$) trajectories cannot remain circular below that radius. The PFDM parameter shifts this divergence: negative $\lambda$ generally pushes the divergence to larger radii (larger ISCO), while positive $\lambda$ pulls it inward, consistent with the potential reshaping shown previously. 

In Fig. \ref{fig:specific_quantities}(b), $\mathcal{E}_{\rm sp}\to \pm\,1$ as $r\to \infty$, as expected for particles at rest at infinity. As $r$ decreases, $\mathcal{E}_{\rm sp}$ drops below unity, indicating bound motion; the quantity $1-\mathcal{E}_{\rm sp}$ is the binding energy per unit mass, which sets a leading estimate for radiative efficiency in thin-disk accretion models. The location where $\mathcal{E}_{\rm sp}$ reaches its minimum while still defined corresponds to the ISCO, so Fig.~\ref{fig:specific_quantities}(b) also provides a practical way to read off how PFDM alters the maximum accretion efficiency.

A stable circular orbit corresponds to a minimum of the effective potential, while an unstable orbit is associated with a maximum. In Newtonian gravity, the effective potential always exhibits a minimum for any value of the angular momentum; consequently, there is no well-defined innermost stable circular orbit (ISCO) with a finite radius.

For circular orbits to be stable, the following conditions must be satisfied:
\begin{equation}
U_\text{eff}(r) = \mathcal{E}^2, \quad \frac{\partial U_\text{eff}(r)}{\partial r} = 0, \quad \frac{\partial^2 U_\text{eff}(r)}{\partial r^2} \geq 0.\label{cc9}
\end{equation}

For marginally stable circular orbits located at radii larger than the black hole shadow radius, we impose the condition
$\frac{\partial^2 U_\text{eff}(r)}{\partial r^2}=0$. Using the effective potential in (\ref{cc5}), we arrive the marginal-stability condition in compact form:
\begin{equation}
    f(r)\,f''(r)-2\,(f'(r))^2+\frac{3}{r}\,f(r)\,f'(r)=0.\label{cc10}
\end{equation}
As a consistency check, for the Schwarzschild metric $f(r)=1-2M/r$ Eq.~\eqref{cc10} reduces to $2M(r-6M)/r^4=0$, recovering the well-known ISCO at $r=6M$.

Substituting the metric function $f(r)$ into Eq.~\eqref{cc10} and expanding yields
\begin{align}
&\frac{2M}{r^3}
-\frac{\lambda}{r^3}\ln\!\frac{r}{|\lambda|}
-\frac{12M^2}{r^4}
+\frac{4M\lambda}{r^4}\left(3\ln\!\frac{r}{|\lambda|}-2\right)
-\frac{\lambda^2}{r^4}
\left(3\ln^2\!\frac{r}{|\lambda|}-4\ln\!\frac{r}{|\lambda|}+2\right) \nonumber\\[4pt]
&+\frac{18\,\eta\, e^{-\gamma}\,M\,Q^2}{r^5}
+\frac{\eta\, e^{-\gamma}\,Q^2\,\lambda}{r^5}
\left(8-9\ln\!\frac{r}{|\lambda|}\right)
-\frac{8\,\eta^2\, e^{-2\gamma}\,Q^4}{r^6}=0.\label{cc11}
\end{align}

In the limit $\lambda=0$, corresponding to the absence of PFDM, the above polynomial equation reduces to a cubic equation as follows:
\begin{equation}
r^3-6 M r^2+9 \eta\,e^{-\gamma} Q^2 r-4 \eta^2 e^{-2\gamma} Q^4/M=0\label{ISCO-condition}
\end{equation}
which further reduces to the Schwarzschild result provided $Q=0$.

The solution of the above Eq. (\ref{ISCO-condition}) is given by
\begin{equation}
r_{\rm ISCO}=2M
+\sqrt[3]{- 9M\eta e^{-\gamma}Q^2+8M^3
+\frac{2\eta^2 e^{-2\gamma}Q^4}{M}+\sqrt{\Delta}}
+\sqrt[3]{- 9M\eta e^{-\gamma}Q^2+8M^3
+\frac{2\eta^2 e^{-2\gamma}Q^4}{M}-\sqrt{\Delta}},\label{ISCO-condition2}
\end{equation}
where the discriminant is as follows:
\begin{align}
\Delta&=\left(9M\eta e^{-\gamma}Q^2-8M^3
-\frac{2\eta^2 e^{-2\gamma}Q^4}{M}\right)^2+\left(3\eta e^{-\gamma}Q^2-4M^2\right)^3\nonumber\\
&=\frac{4\,\eta^{2} e^{-2\gamma} Q^{4}}{M^{2}}
\left(\eta e^{-\gamma} Q^{2}-M^{2}\right)
\left(\eta e^{-\gamma} Q^{2}-\frac{5}{4}M^{2}\right).\label{ISCO-condition3}
\end{align}  
The nature of the roots depends on the discriminant $\Delta$:
\begin{itemize}
\item If $\Delta>0$, the cubic equation has one real root (given above) and two complex conjugate roots.
\item If $\Delta=0$, the cubic equation has multiple real roots (at least two coincide).
\item If $\Delta<0$, the cubic equation has three distinct real roots, which are most conveniently expressed in trigonometric form.
\end{itemize}

\begin{itemize}
    \item For $\eta=+1$, corresponding to ModMax black hole, the discriminant simplifies as
    \begin{equation}
    \Delta=\frac{4 e^{-2\gamma} Q^{4}}{M^{2}}
\left(e^{-\gamma} Q^{2}-M^{2}\right)
\left(e^{-\gamma} Q^{2}-\frac{5}{4}M^{2}\right)>0
    \end{equation}
    provided $M^2 < 0.8 e^{-\gamma} Q^2$.
    \item For $\eta=-1$, corresponding to phantom ModMax black hole, the discriminant simplifies as
    \begin{equation}
        \Delta=\frac{4 e^{-2\gamma} Q^{4}}{M^{2}}
\left(e^{-\gamma} Q^{2}+M^{2}\right)
\left(e^{-\gamma} Q^{2}+\frac{5}{4}M^{2}\right)>0.
    \end{equation}
\end{itemize}

\begin{figure}[tbhp]
\centering
\includegraphics[width=\linewidth]{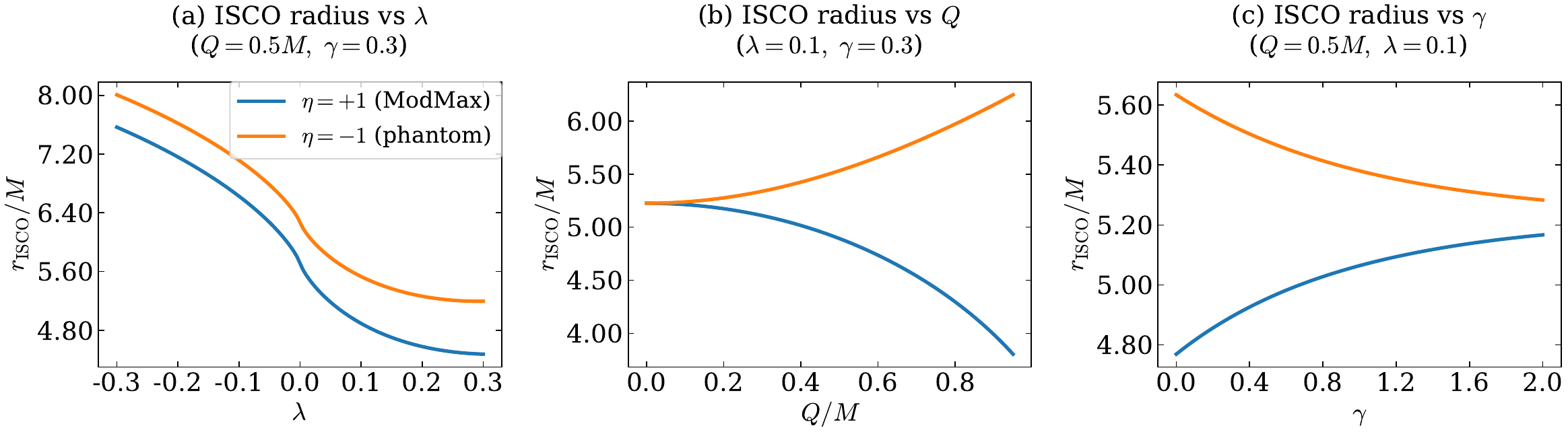}
\caption{ISCO radius for timelike circular orbits around the Mod(A)Max black hole with PFDM. Panel (a) shows $r_{\rm ISCO}/M$ as a function of the PFDM parameter $\lambda$ at fixed $(Q,\gamma)$; panel (b) shows $r_{\rm ISCO}/M$ as a function of the charge-to-mass ratio $Q/M$ at fixed $(\lambda,\gamma)$;
panel (c) shows $r_{\rm ISCO}/M$ as a function of the ModMax parameter $\gamma$ at fixed $(Q,\lambda)$. In all panels, $r_{\rm ISCO}$ is obtained numerically from the marginal-stability condition $f\,f''-2\,(f')^2+\frac{3}{r}\,f\,f'=0$, selecting the smallest physical root outside the event horizon. Curves are shown for $\eta=+1$ (ModMax) and $\eta=-1$ (phantom).}
\label{fig:isco_radius}
\end{figure}

The exact analytical solution of the above equation will give us the ISCO radius $r_{\rm ISCO}$. Note that the exact analytical solution is quite challenging due to the presence of the logarithmic function. However, one can determine the numerical results by selecting suitable values of parameters. Thereby, it becomes evident that the ISCO radius depends on the electric charge $Q$, the PFDM parameter $\lambda$, and the ModMax parameter \(\gamma\). 

Using Eq.~\eqref{cc11}, we compute the ISCO radius by (i) locating the outer horizon $r_h$ from $f(r_h)=0$ and (ii) selecting the smallest root of Eq.~\eqref{cc11} in the domain $r>r_h$ (where $f>0$). Figure~\ref{fig:isco_radius} summarizes how $r_{\rm ISCO}$ shifts when varying $\lambda$, $Q/M$, and $\gamma$ for both branches $\eta=\pm 1$.

\section{Thermodynamics}

In this section, the thermodynamic properties of the black hole are examined by analyzing its temperature, Gibbs free energy, and heat capacity. These quantities provide insights into the stability and phase structure of the black hole. In particular, the behavior of the heat capacity allows us to identify regions of thermodynamic stability and possible phase transitions, while the Gibbs free energy offers a global view of the preferred thermodynamic phases. The influence of the geometric parameters and the surrounding perfect fluid dark matter on these thermodynamic quantities is also investigated, revealing how modifications to the spacetime geometry affect the black hole's thermal behavior.

The mass of the black hole can be determined using the condition $f(r=r_h)=0$, where $r_h$ is the event horizon radius. In our case at hand, we find
\begin{equation}
    M=\frac{r_h}{2}\left[1+\eta\,e^{-\gamma}\,\frac{Q^2}{r^2_h}+\frac{\lambda}{r_h}\ln\!\frac{r_h}{|\lambda|}\right].\label{dd1}
\end{equation}

\begin{itemize}
    \item For $\eta=+1$, corresponding to ModMax black hole, the black hole mass simplifies as
    \begin{equation}
    M=\frac{r_h}{2}\left[1+e^{-\gamma}\,\frac{Q^2}{r^2_h}+\frac{\lambda}{r_h}\ln\!\frac{r_h}{|\lambda|}\right].\label{dd1a}
\end{equation}
    \item For $\eta=-1$, corresponding to phantom ModMax black hole, the black hole mass simplifies as
    \begin{equation}
    M=\frac{r_h}{2}\left[1-e^{-\gamma}\,\frac{Q^2}{r^2_h}+\frac{\lambda}{r_h}\ln\!\frac{r_h}{|\lambda|}\right].\label{dd1b}
\end{equation}
\end{itemize}

\begin{figure}[tbhp]
    \centering
    \includegraphics[width=\textwidth]{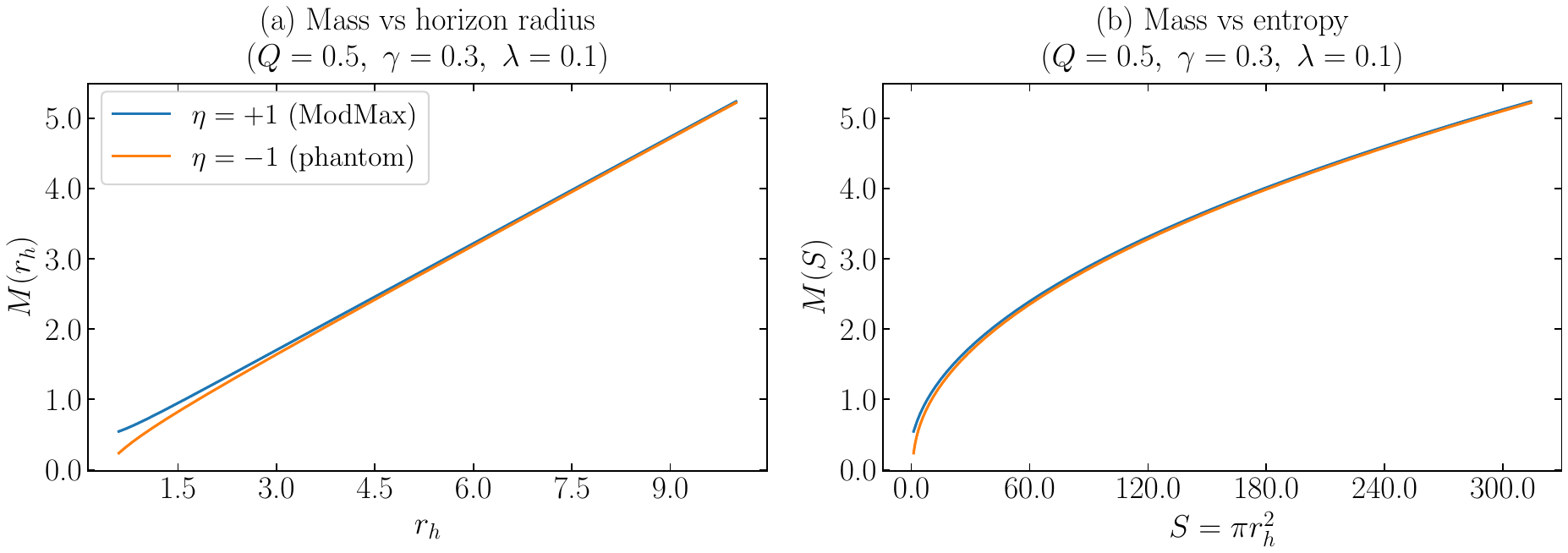}
    \caption{Thermodynamic mass of the spherically symmetric Mod(A)Max black hole surrounded by perfect-fluid dark matter (PFDM), shown as a function of the horizon radius and of the Bekenstein--Hawking entropy. Panel (a) displays the mass function $M(r_h)$ obtained from the horizon condition $f(r_h)=0$, for fixed $(Q,\gamma,\lambda)=(0.5,\,0.3,\,0.1)$ and for both choices of the electromagnetic-sector sign $\eta=\pm1$. Panel (b) shows the same relation reparametrized in terms of the entropy $S=\pi r_h^2$, i.e., $M(S)=M\!\left(r_h=\sqrt{S/\pi}\right)$. The phantom case ($\eta=-1$) yields systematically smaller masses for a given horizon size (or entropy), reflecting the sign reversal of the Mod(A)Max contribution in the lapse function.}
    \label{fig:thermo_mass}
\end{figure}
Figure~\ref{fig:thermo_mass} summarizes the algebraic relation between the ADM mass parameter and the horizon location implied by the metric function in Eq.~(\ref{function}). Imposing the horizon condition $f(r_h)=0$ allows one to express the mass in the form (\ref{dd1}), which is plotted in Fig.~\ref{fig:thermo_mass}(a) for representative parameters. This curve provides an immediate consistency check for the parameter space: for any chosen $M$, the intersections with the horizontal line $M=\mathrm{const.}$ determine the possible horizon radii (and hence whether multiple horizons can occur). Rewriting the same relation in terms of the Bekenstein-Hawking entropy \cite{Bekenstein1973, Bekenstein1974,Hawking1975}, \(S=\pi r_h^2,\) 
yields $M(S)$ in Fig.~\ref{fig:thermo_mass}(b), which is the natural form for thermodynamic analyses (e.g., when constructing $T(S)$, heat capacities, or free energies). For fixed $(Q,\gamma,\lambda)$, the phantom branch ($\eta=-1$) lies below the ModMax branch ($\eta=+1$), indicating that the sign flip in the electromagnetic contribution reduces the mass required to support a given horizon size (or entropy), and therefore shifts the thermodynamic scales of the solution.

As the selected lapse function of the black hole metric is asymptotically flat and due to the spherical symmetry space-time, the surface gravity is given by (in the signature $+2$) \cite{Hawking1975,GibbonsHawking1977,BardeenCarterHawking1973}
\begin{equation}
    \kappa=-\frac{1}{2} \lim_{r \to r_h} \frac{\partial_r g_{tt}}{\sqrt{-g_{tt}\,g_{rr}}}=\frac{f'(r_h)}{2}.\label{dd2}
\end{equation}
Therefore, the Hawking temperature is given by
\begin{equation}
    T=\frac{\kappa}{2\pi}=\frac{1}{4\pi r_h}\left(1-\eta\,\frac{e^{-\gamma}Q^2}{r_h^2}+\frac{\lambda}{r_h}\right).\label{dd3}
\end{equation}

\begin{itemize}
    \item For $\eta=+1$, corresponding to ModMax black hole, the Hawking temperature simplifies as
    \begin{equation}
    T=\frac{1}{4\pi r_h}\left(1-\frac{e^{-\gamma}Q^2}{r_h^2}+\frac{\lambda}{r_h}\right).\label{dd3a}
\end{equation}
\item For $\eta=-1$, corresponding to phantom ModMax black hole, the Hawking temperature simplifies as
    \begin{equation}
    T=\frac{1}{4\pi r_h}\left(1+\frac{e^{-\gamma}Q^2}{r_h^2}+\frac{\lambda}{r_h}\right).\label{dd3b}
\end{equation}
    
\end{itemize}

\begin{figure}[htbp]
    \centering
    \includegraphics[width=\textwidth]{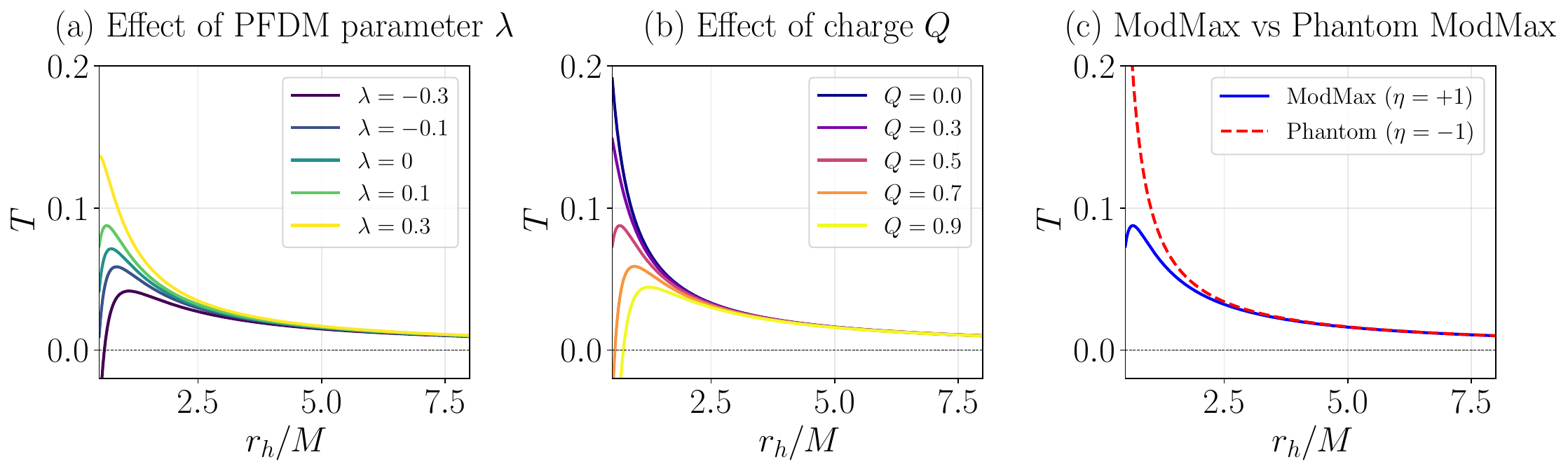}
    \caption{The Hawking temperature $T$ as a function of the horizon radius $r_h/M$ for the Mod(A)Max black hole surrounded by perfect fluid dark matter. Panel~(a) shows the effect of the PFDM parameter $\lambda$: negative values of $\lambda$ increase the temperature at small horizon radii, while positive values suppress it. The temperature approaches zero at a critical horizon radius that depends on $\lambda$, indicating the existence of extremal black hole configurations. Panel~(b) illustrates the influence of the electric charge $Q$; increasing charge reduces the temperature and shifts the zero-temperature point to larger horizon radii, consistent with the approach to extremality in charged black hole solutions. Panel~(c) compares the ModMax ($\eta = +1$) and phantom ModMax ($\eta = -1$) cases, revealing that the phantom configuration exhibits higher temperatures at small radii due to the sign reversal in the electromagnetic contribution to the surface gravity. The horizontal dashed line at $T = 0$ marks the extremal limit where the black hole ceases to emit Hawking radiation.}
    \label{fig:hawking_temp}
\end{figure}

The behavior of the Hawking temperature as a function of the horizon radius is illustrated in Fig.~\ref{fig:hawking_temp}. This figure provides crucial insights into the thermal properties of the black hole and the conditions under which it can achieve thermodynamic equilibrium.

Figure~\ref{fig:hawking_temp} should be interpreted as follows. For a Schwarzschild black hole, the temperature scales as $T \propto 1/r_h$, monotonically decreasing with increasing horizon size. The introduction of charge, PFDM, and the ModMax parameter modifies this behavior significantly. In Fig~\ref{fig:hawking_temp}(a), we observe that the PFDM parameter $\lambda$ acts as a regulator of the thermal emission: negative $\lambda$ enhances the temperature gradient, making smaller black holes hotter, while positive $\lambda$ tends to flatten the temperature profile. This has direct implications for the evaporation rate, since the Stefan-Boltzmann law relates the power output to $T^4$.

Figure~\ref{fig:hawking_temp}(b) demonstrates the well-known effect of charge on black hole thermodynamics. As $Q$ increases, the temperature decreases for a given horizon radius, and the black hole approaches an extremal state where $T \to 0$. This behavior is reminiscent of Reissner-Nordstr\"om black holes, but here modified by the ModMax nonlinear electrodynamics through the factor $e^{-\gamma}$.

The comparison in Fig.~\ref{fig:hawking_temp}(c) between ModMax and phantom configurations is particularly instructive. The sign flip $\eta \to -\eta$ effectively reverses the role of the electromagnetic contribution in the surface gravity calculation. For the phantom case ($\eta = -1$), the charge term in Eq.~(\ref{dd3}) adds constructively to the temperature rather than subtracting, leading to higher temperatures and fundamentally different evaporation dynamics.

The entropy of the thermodynamic system is given by \cite{Bekenstein1973, Bekenstein1974,Hawking1975}
\begin{equation}
    S=\int \frac{dM}{T}.\label{dd4}
\end{equation}
Using the mass of the back hole given in (\ref{dd1}) and temperature in (\ref{dd3}), we find
\begin{equation}
S=\int 2\pi r_h dr_h=\pi r^2_h=\mathcal{A}/4 \label{dd5}
\end{equation}
which is the Bekenstein-Hawking entropy equal to a quarter of the horizon area.

The specific heat capacity of the thermodynamic system is given by \cite{Hawking1975,Davies1977}
\begin{equation}
C=\frac{dM}{dT}=-2 \pi r_h^2 \,\left(\frac{r_h^2 -\eta e^{-\gamma} Q^2+\lambda\,r_h}{r_h^2 -3 \eta e^{-\gamma} Q^2+2 \lambda\,r_h}\right).\label{dd6}
\end{equation}

\begin{itemize}
    \item For $\eta=+1$, corresponding to ModMax black hole, the specific heat simplifies as
    \begin{equation}
    C=-2 \pi r_h^2 \,\left(\frac{r_h^2 -e^{-\gamma} Q^2+\lambda\,r_h}{r_h^2 -3e^{-\gamma} Q^2+2 \lambda\,r_h}\right).\label{dd6a}
\end{equation}
    \item For $\eta=-1$, corresponding to phantom ModMax black hole, the specific heat simplifies as
    \begin{equation}
    C=-2 \pi r_h^2 \,\left(\frac{r_h^2 +e^{-\gamma} Q^2+\lambda\,r_h}{r_h^2 +3e^{-\gamma} Q^2+2 \lambda\,r_h}\right).\label{dd6b}
\end{equation}
\end{itemize}

\begin{figure}[ht!]
    \centering
    \includegraphics[width=\textwidth]{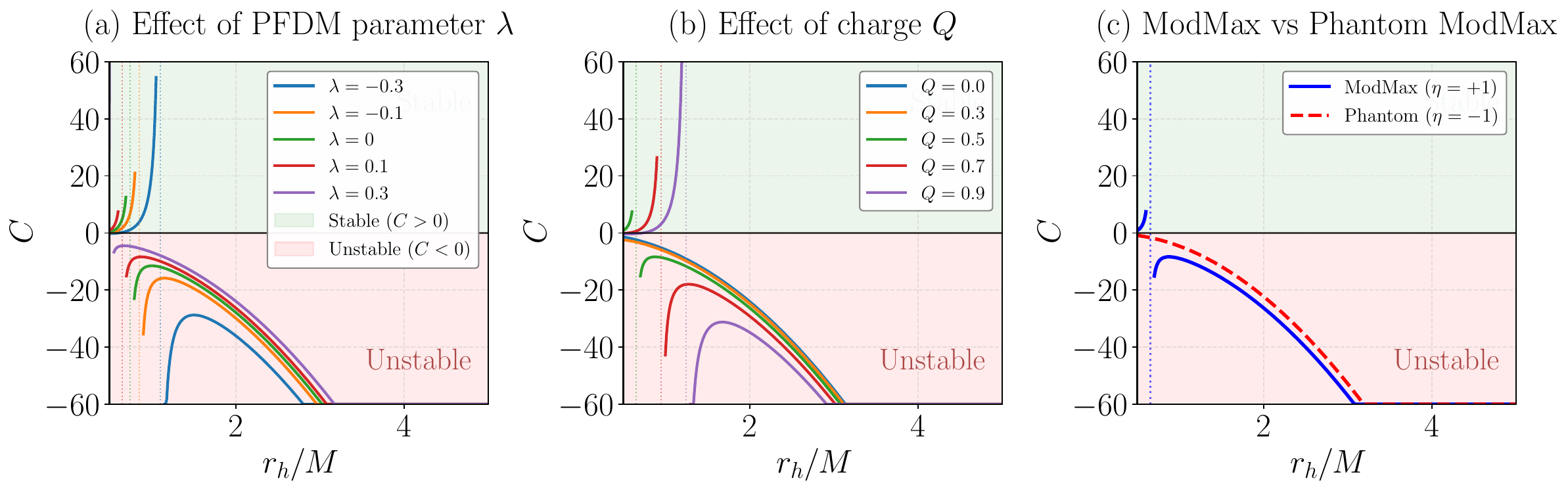}
    \caption{The heat capacity $C$ as a function of the horizon radius $r_h/M$ for the Mod(A)Max black hole surrounded by perfect fluid dark matter. Panel~(a) demonstrates the effect of the PFDM parameter $\lambda$: the divergence point, which signals a phase transition, shifts to different horizon radii depending on $\lambda$. Negative $\lambda$ values push the divergence to larger radii, while positive values bring it closer to the horizon. Panel~(b) shows the influence of the electric charge $Q$; increasing charge moves the phase transition point and modifies the stability regions. For small black holes (left of the divergence), $C < 0$ indicates thermodynamic instability, while for large black holes (right of the divergence), $C < 0$ persists in certain parameter regimes. Panel~(c) compares ModMax ($\eta = +1$) and phantom ($\eta = -1$) configurations, showing that the phantom case exhibits a shifted divergence structure due to the modified electromagnetic contribution. The horizontal dashed line at $C = 0$ separates stable ($C > 0$) from unstable ($C < 0$) configurations.}
    \label{fig:heat_capacity}
\end{figure}

The sign of $C$ determines the local thermodynamic stability: $C> 0$ indicates a stable configuration, while $C< 0$ signals thermodynamic instability. Phase transitions occur at points where $C$ diverges, corresponding to critical points in the black hole's thermodynamic phase space. The behavior of the heat capacity is shown in Fig.~\ref{fig:heat_capacity}.

The heat capacity analysis in Fig.~\ref{fig:heat_capacity} reveals the rich phase structure of the Mod(A)Max black hole with PFDM. The divergences in $C$ occur when the denominator in Eq.~(\ref{dd6}) vanishes, i.e., when
\begin{equation}
r_h^2 - 3\eta e^{-\gamma} Q^2 + 2\lambda r_h = 0\Rightarrow r_h=-\lambda+\sqrt{\lambda^2+3 \eta e^{-\gamma} Q^2}.
\end{equation}
This condition defines a critical horizon radius $r_h^{\rm crit}$ that marks the boundary between thermodynamically stable and unstable phases provided $\eta=+1$ otherwise not.

In standard black hole thermodynamics, a negative heat capacity implies that as the black hole loses energy through Hawking radiation, its temperature increases, leading to runaway evaporation. Conversely, a positive heat capacity allows for thermal equilibrium with a heat bath. The presence of PFDM and the ModMax parameter introduces additional control over this phase structure, potentially stabilizing configurations that would otherwise be unstable.

The divergence in $C$ signals a second-order phase transition in the canonical ensemble. Near this critical point, the black hole exhibits enhanced fluctuations and the system may transition between small and large black hole branches. The detailed analysis of these phase transitions requires examination of the free energy landscape, which we address in the following section on thermodynamic topology.

Next, we evaluate the Gibbs free energy of the thermodynamic system. Considering the black hole mass as the internal energy of the system, the Gibbs free energy is given by
\begin{equation}
F=M-T\,S=\frac{r_h}{4}\left[1+3\,\eta e^{-\gamma}\frac{Q^2}{r_h^2}+2\,\frac{\lambda}{r_h}\ln\!\frac{r_h}{|\lambda|}-\frac{\lambda}{r_h}\right].\label{gibbs}
\end{equation}
\begin{itemize}
    \item For $\eta=+1$, corresponding to ModMax black hole, the Gibbs free energy simplifies as
    \begin{equation}
    F=\frac{r_h}{4}\left[1+3 e^{-\gamma}\frac{Q^2}{r_h^2}+2\,\frac{\lambda}{r_h}\ln\!\frac{r_h}{|\lambda|}-\frac{\lambda}{r_h}\right].\label{gibbsa}
\end{equation}
    \item For $\eta=-1$, corresponding to phantom ModMax black hole, the Gibbs free energy simplifies as
    \begin{equation}
    F=\frac{r_h}{4}\left[1-3 e^{-\gamma}\frac{Q^2}{r_h^2}+2\,\frac{\lambda}{r_h}\ln\!\frac{r_h}{|\lambda|}-\frac{\lambda}{r_h}\right].\label{gibbsb}
\end{equation}
\end{itemize}

Finally, we discuss the modified Smarr formula for the thermodynamic system. Noted that in addition to the entropy $S$ and the electric charge $Q$ as variables, PFDM parameter $\lambda$ is assumed as thermodynamic intensive variable. The black hole mass $M$ given in Eq.~(\ref{dd1}) in terms of entropy $S$ can be expressed as
\begin{equation}
M=\frac{1}{2}\sqrt{\frac{S}{\pi}}\left[1+\pi\,\eta\,e^{-\gamma}\,\frac{Q^2}{S}+\lambda\,\sqrt{\frac{\pi}{S}}\ln\!\left(\frac{1}{|\lambda|}\,\sqrt{\frac{S}{\pi}}\right)\right].\label{dd7}
\end{equation}

Thereby, one can see that the internal energy of mass mass $M=M(S, Q, \lambda)$, and hence, the differential mass can be written as
\begin{equation}
    dM=T\,dS+\Phi\,dQ+\Phi_{\lambda} d\lambda,\label{dd8}
\end{equation}
where $T=\left(\frac{\partial M}{\partial S}\right)_{Q, \lambda}$ is the temperature and the potentials $\Phi$ and $\Phi_{\lambda}$ are given by
\begin{align}
    &\Phi=\left(\frac{\partial M}{\partial Q}\right)_{S, \lambda}=\eta\,e^{-\gamma}\,\sqrt{\frac{\pi}{S}}\,Q=\eta\,e^{-\gamma}\,\frac{Q}{r_h}.\label{dd9}\\
    &\Phi_{\lambda}=\left(\frac{\partial M}{\partial \lambda}\right)_{S, Q}=\frac{1}{2}\ln\!\left(\frac{1}{|\lambda|}\,\sqrt{\frac{S}{\pi}}\right)-\frac{1}{2}=\frac{1}{2} \left( \ln\!\frac{r_h}{|\lambda|} - 1 \right).\label{dd10}
\end{align}

Now, we need to evaluate the quantity $(2 T S+\Phi\,Q+\Phi_{\lambda}\,\lambda)$ using the above results. We find  that
\begin{equation}
2 T S+\Phi\,Q+\Phi_{\lambda}\,\lambda=\frac{r_h}{2} + \frac{\eta e^{-\gamma} Q^2}{2 r_h} + \frac{\lambda}{2} \ln\frac{r_h}{|\lambda|} = M\label{d11}
\end{equation}
which is equal to the internal energy of the thermodynamic system, and thus, obeying the Smarr formula, confirming the consistency of our thermodynamic framework.

\section{Thermodynamic Topology}\label{sec5}

In the previous section, we analyzed the thermodynamic properties of the Mod(A)Max black hole surrounded by perfect fluid dark matter, deriving key quantities such as the Hawking temperature and the specific heat capacity. While these quantities provide valuable insight into local thermodynamic stability, the global phase structure of the system can be more transparently understood from a topological perspective.

Recently, Wei \emph{et al.}~\cite{Wei2022a,Wei2022b} proposed a novel thermodynamic--topology correspondence, in which black hole configurations are interpreted as topological defects in an auxiliary parameter space. Within this framework, the thermodynamic stability and phase transitions of black holes are characterized by topological invariants, such as winding numbers. This approach has been successfully applied to a wide variety of black hole solutions in general relativity and modified gravity theories, revealing that seemingly diverse phase structures can be classified into a small number of universal topological classes.

In this section, we adopt the generalized Gibbs free energy formalism developed by Wei \emph{et al.}~\cite{Wei2022a,Wei2022b} to investigate the thermodynamic topology of the selected black hole solution. Our primary focus is on elucidating how the presence of the dark matter parameter and ModMax parameter influence the topological properties and phase structure of the system. Owing to the equivalence between the black hole mass and internal energy, the generalized Gibbs free energy can be formulated as a standard thermodynamic potential~\cite{Wei2022a,Wei2022b},
\begin{equation}
\mathcal{F}=M(r_{h})-\frac{S}{\tau}, \label{ff1}
\end{equation}
where $\tau$ denotes the Euclidean time (interpreted as the inverse temperature at the zero points) outside the horizon, and $S=\pi r_{h}^{2}$ is the entropy of the black hole derived earlier. The generalized free energy becomes an on-shell quantity only when the condition $\tau = 1/T$ is satisfied. In our case at hand, we have this generalized energy function
\begin{equation}
  \mathcal{F}=\frac{r_h}{2}\left[1+\eta\,e^{-\gamma}\,\frac{Q^2}{r^2_h}+\frac{\lambda}{r_h}\ln\!\frac{r_h}{|\lambda|}\right]-\pi r^2_h/\tau.\label{ff2}
\end{equation}

\begin{figure}[htbp]
    \centering
    \includegraphics[width=\textwidth]{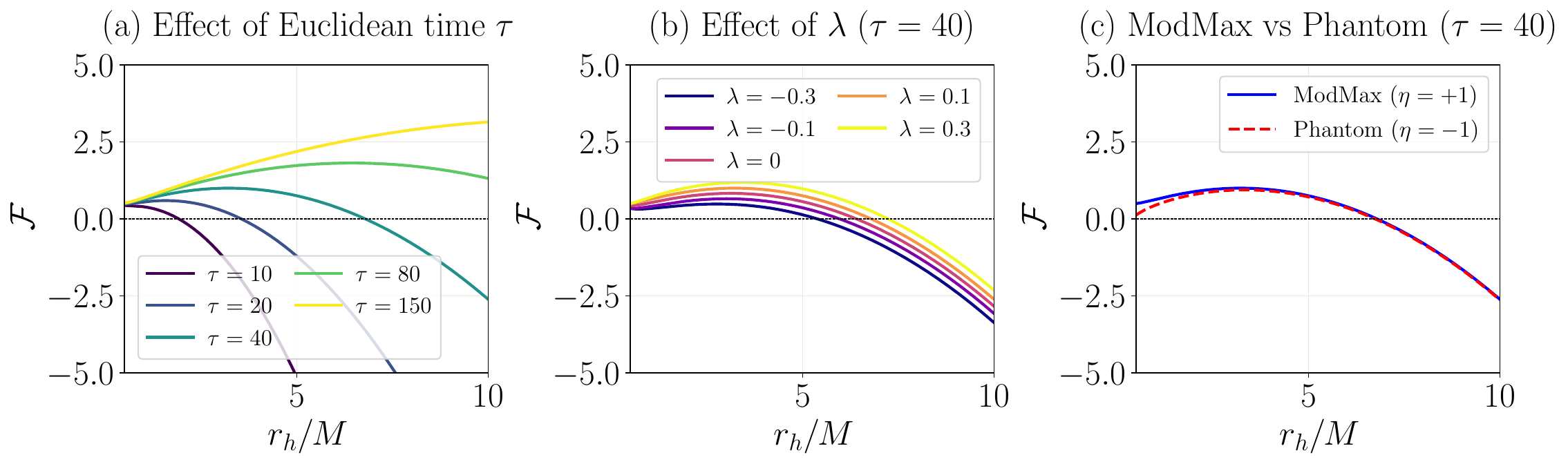}
    \caption{The generalized Gibbs free energy $\mathcal{F}$ as a function of the horizon radius $r_h/M$ for the Mod(A)Max black hole surrounded by perfect fluid dark matter. Panel~(a) illustrates the effect of the Euclidean time $\tau$: for small $\tau$ (high temperature), the free energy is dominated by the entropy term and decreases monotonically, while for large $\tau$ (low temperature), a local minimum develops indicating a stable black hole phase. The transition between these regimes signals the Hawking-Page-like phase transition. Panel~(b) shows the influence of the PFDM parameter $\lambda$ at fixed $\tau = 40$; negative $\lambda$ values deepen the free energy minimum and shift it to larger radii, enhancing the stability of the large black hole phase. Panel~(c) compares ModMax ($\eta = +1$) and phantom ($\eta = -1$) configurations, demonstrating that the phantom case exhibits a shallower free energy minimum located at larger horizon radii. The horizontal dashed line at $\mathcal{F} = 0$ marks the threshold for thermodynamic preference; phases with $\mathcal{F} < 0$ are globally preferred over thermal radiation.}
    \label{fig:gibbs_energy}
\end{figure}

The behavior of the generalized Gibbs free energy is illustrated in Fig.~\ref{fig:gibbs_energy}, which shows how the free energy landscape depends on the model parameters.

The free energy analysis in Fig.~\ref{fig:gibbs_energy} provides a global view of the thermodynamic phase structure. The competition between the mass term (which increases with $r_h$) and the entropy term (which grows as $r_h^2/\tau$) determines the shape of the free energy landscape. For small $\tau$ (equivalently, high temperature), the entropy term dominates and $\mathcal{F}$ decreases monotonically, indicating that larger black holes are always preferred. As $\tau$ increases (temperature decreases), a local minimum develops, signaling the emergence of a stable black hole phase that coexists with thermal radiation.

The PFDM parameter $\lambda$ plays a crucial role in modifying this landscape. As shown in panel~(b), negative $\lambda$ values enhance the depth of the free energy minimum, making the black hole phase more thermodynamically stable. This can be understood from the logarithmic contribution to the mass in Eq.~(\ref{dd1}), which modifies the balance between gravitational binding energy and thermal effects.

The components of the vector field $\boldsymbol{\phi}_{\mathcal{F}}$ associated with the generalized free energy are defined as~\cite{Wei2022a,Wei2022b}
\begin{align}
\phi_{\mathcal{F}}^{r_{h}}&= \partial_{r_{h}}\mathcal{F}=\frac{1}{2} \left(1
-\eta\,\frac{e^{-\gamma}Q^2}{r_h^2}
+\frac{\lambda}{r_h}\right)-2\pi r_h/\tau,\label{ff3}\\
    \phi_{\mathcal{F}}^{\theta}&= -\cot\theta\,\csc\theta. \label{ff3b}
\end{align}
Its magnitude is normalized through
\begin{equation}
\|\phi_{\mathcal{F}}\| = \sqrt{ \left( \phi_{\mathcal{F}}^{r_{h}} \right)^{2} + \left( \phi_{\mathcal{F}}^{\theta} \right)^{2} }=\sqrt{\left[\frac{1}{2} \left(1
-\eta\,\frac{e^{-\gamma}Q^2}{r_h^2}
+\frac{\lambda}{r_h}\right)-2\pi r_h/\tau\right]^2+\frac{\cot^2 \theta}{\sin^2 \theta}}.\label{ff4}
\end{equation}
Thereby, the normalized unit vector field ${\bf n}_{\mathcal{F}}$ is given by
\begin{align}
&n_{\mathcal{F}}^{r_{h}} = \frac{\phi^{\mathcal{F}}_{r_{h}}}{\|\phi_{\mathcal{F}}\|}=\frac{\frac{1}{2} \left(1-\eta\,\frac{e^{-\gamma}Q^2}{r_h^2}
+\frac{\lambda}{r_h}\right)-2\pi r_h/\tau}{\sqrt{\left[\frac{1}{2} \left(1-\eta\,\frac{e^{-\gamma}Q^2}{r_h^2}+\frac{\lambda}{r_h}\right)-2\pi r_h/\tau\right]^2+\frac{\cot^2 \theta}{\sin^2 \theta}}},\label{ff5}\\ 
&n_{\mathcal{F}}^{\theta} = \frac{\phi^{\mathcal{F}}_{\theta}}{\|\phi_{\mathcal{F}}\|}=-\frac{\cot\theta\,\csc\theta}{\sqrt{\left[\frac{1}{2} \left(1-\eta\,\frac{e^{-\gamma}Q^2}{r_h^2}
+\frac{\lambda}{r_h}\right)-2\pi r_h/\tau\right]^2+\frac{\cot^2 \theta}{\sin^2 \theta}}}. \label{ff5b}
\end{align}

\begin{figure}[htbp]
    \centering
    \includegraphics[width=\textwidth]{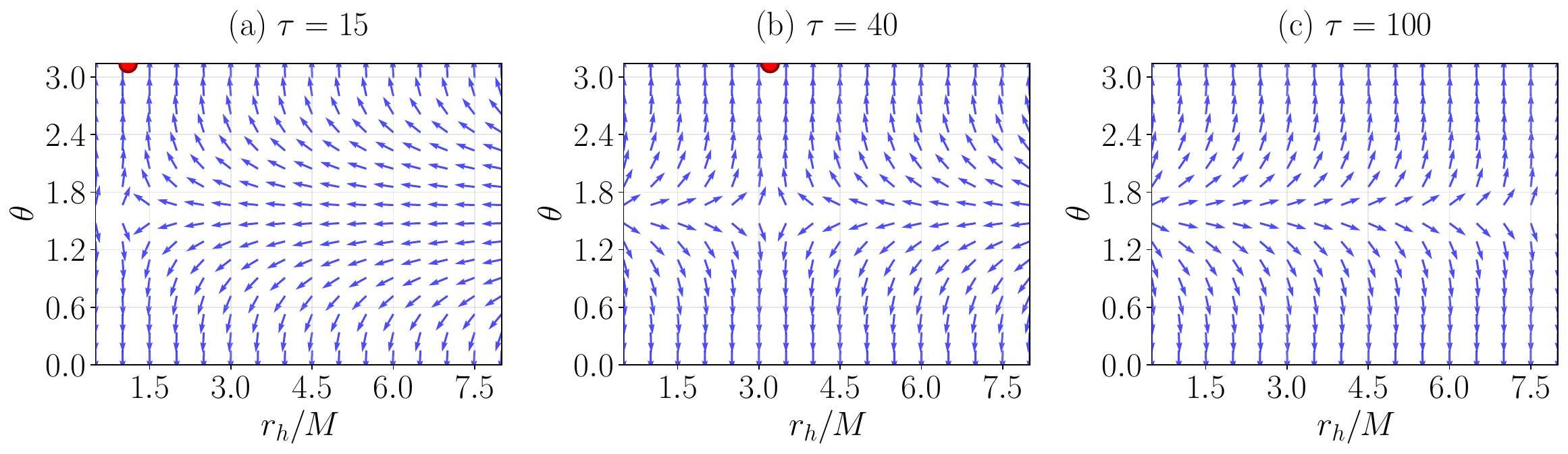}
    \caption{The normalized unit vector field $\mathbf{n}_{\mathcal{F}}$ in the $(r_h, \theta)$ parameter space for the Mod(A)Max black hole with PFDM, shown for different values of the Euclidean time $\tau$. Panel~(a) corresponds to $\tau = 15$ (high temperature), panel~(b) to $\tau = 40$ (intermediate temperature), and panel~(c) to $\tau = 100$ (low temperature). The vector field points outward at the boundaries $\theta = 0$ and $\theta = \pi$ due to the divergence of $\phi_{\mathcal{F}}^{\theta}$. The red circles mark the zero points of the vector field, located at $\theta = \pi$ and the horizon radius satisfying $\partial_{r_h}\mathcal{F} = 0$. Each zero point represents a distinct thermodynamic phase, and its winding number characterizes the local stability: positive winding indicates a stable phase, while negative winding signals instability. The parameters are set to $Q = 0.5M$, $\gamma = 0.3$, $\lambda = 0.1$, and $\eta = +1$ (ModMax).}
    \label{fig:vector_field}
\end{figure}

The normalized vector field is visualized in Fig.~\ref{fig:vector_field}, which shows the flow structure in the $(r_h, \theta)$ parameter space for different values of the Euclidean time $\tau$.

The topological interpretation of Fig.~\ref{fig:vector_field} is as follows. The zeros of the vector field $\boldsymbol{\phi}_{\mathcal{F}}$ correspond to on-shell black hole configurations where the generalized free energy is stationary. These points act as topological defects in the parameter space, and their nature (stable vs. unstable) is encoded in the winding number computed by integrating the vector field along a closed contour surrounding each zero.

The zeros of the vector field $\boldsymbol{\phi}_{\mathcal{F}}$ are determined by two simultaneous conditions: (i) $\theta=\pi$, and (ii) $\partial_{r_{h}}\mathcal{F}=0$, where the latter corresponds to stationary points of the generalized free energy with respect to the horizon radius. Each zero point represents a distinct thermodynamic phase of the black hole, such as the small or large black hole branches. The associated winding number characterizes the local thermodynamic stability of each phase: a positive winding number indicates a stable configuration, whereas a negative winding number signals an unstable one.

It is worth noting that the component $\phi_{\mathcal{F}}^{\theta}$ diverges at $\theta=0$ and $\theta=\pi$, which causes the vector field to point outward at the boundaries of the parameter space. Throughout this analysis, the variables are taken to lie in the ranges $0 \leq r_{h} < \infty$ and $0 \leq \theta \leq \pi$.

From the above normalized vector field, it becomes evident that the electric charge $Q$, the PFDM parameter $\lambda$, the ModMax parameter \(\gamma\), and the black hole mass \(M\) modifies the vector field of the free-energy. Consequently, thermodynamic topological properties of the black hole depends on these geometric parameters.

Now, we determine Euclidean time $\tau$ at zero points of the vector field $\phi_{\mathcal{F}}$. This can be determined using the condition $\partial_{r_{h}}\mathcal{F}=0$ which yields
\begin{equation}
    \tau=\frac{4 \pi r_h}{1
-\eta\,\frac{e^{-\gamma}Q^2}{r_h^2}
+\frac{\lambda}{r_h}}=\frac{1}{T}.\label{ff6}
\end{equation}

\begin{figure}[htbp]
    \centering
    \includegraphics[width=\textwidth]{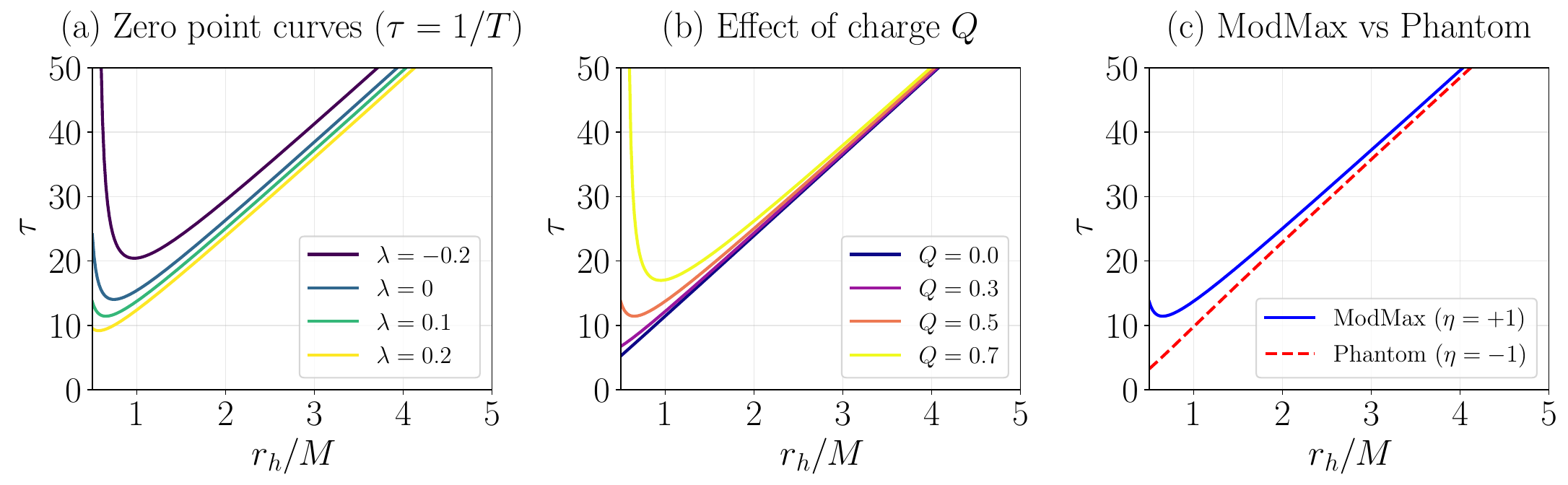}
    \caption{The zero point curves $\tau = 1/T$ in the $(r_h, \tau)$ plane for the Mod(A)Max black hole surrounded by perfect fluid dark matter. Panel~(a) shows the effect of the PFDM parameter $\lambda$: negative values shift the curve upward and to the right, indicating that larger black holes can exist at lower temperatures (larger $\tau$). Positive $\lambda$ values have the opposite effect, compressing the accessible phase space. Panel~(b) illustrates the influence of the electric charge $Q$; increasing charge extends the curve to larger $\tau$ values, reflecting the approach to extremality where $T \to 0$. Panel~(c) compares ModMax ($\eta = +1$) and phantom ($\eta = -1$) configurations, showing that the phantom case exhibits a more extended curve reaching larger horizon radii at fixed $\tau$. These curves define the thermodynamic phase boundaries and determine which black hole configurations are accessible at a given temperature.}
    \label{fig:zero_points}
\end{figure}

The relationship between the horizon radius and the Euclidean time at the zero points, which defines the curve $\tau = 1/T$, is illustrated in Fig.~\ref{fig:zero_points}. This curve represents the locus of on-shell black hole configurations in thermodynamic equilibrium.

The zero point curves in Fig.~\ref{fig:zero_points} encode the complete information about the equilibrium thermodynamics of the system. For a given value of $\tau$ (or equivalently, a fixed temperature of the thermal bath), the intersection with the curve determines the horizon radius of the black hole in thermal equilibrium. Multiple intersections at a single $\tau$ would indicate phase coexistence, while the turning points of the curve mark critical temperatures where phase transitions occur.

The modification of these curves by the PFDM parameter $\lambda$ demonstrates that dark matter can significantly alter the thermodynamic phase diagram. In particular, negative $\lambda$ extends the range of accessible equilibrium configurations, potentially stabilizing black holes that would otherwise be thermodynamically forbidden.

\section{Scalar Perturbations: The Massless Klein-Gordon Equation}

Scalar perturbations provide an effective probe of the stability and dynamical properties of black hole spacetimes. In the presence of perfect fluid dark matter and nonlinear electrodynamics, the background geometry deviates from the standard Schwarzschild configuration, thereby modifying the evolution of massive and massless scalar fields governed by the Klein-Gordon equation. After separating the variables, the perturbation equation reduces to a Schr\"odinger-like wave equation with an effective potential that explicitly depends on the Mod(A)Max and PFDM parameters, which in turn determines the quasinormal mode (QNM) spectrum \cite{SF1,SF2,SF3}. The combined effects of PFDM and nonlinear electrodynamics enrich the perturbative dynamics of the black hole and may leave observable imprints in the ringdown phase of gravitational waves. Consequently, QNM analyses-using techniques such as the WKB approximation and time-domain evolution-serve as powerful tools for probing exotic matter distributions in astrophysical environments \cite{SF1,SF3}.

The dynamics of a massless scalar field $\Psi$ propagating in the black hole background is governed by the Klein--Gordon equation
\begin{equation}
\frac{1}{\sqrt{-g}}\partial_\mu\left(\sqrt{-g}g^{\mu\nu}\partial_\nu\Psi\right) = 0,
\label{sf1}
\end{equation}
where $g_{\mu\nu}$ denotes the spacetime metric.

We adopt the following ansatz for the scalar field:
\begin{equation}
\Psi = \frac{\psi(r)}{r}Y^{m}_{\ell}(\theta,\phi)e^{-i\omega t},
\label{sf2}
\end{equation}
where $\omega$ is the oscillation frequency, $Y^m_\ell(\theta,\phi)$ are the spherical harmonics with multipole number $\ell$ and azimuthal number $m$, and $\psi(r)$ represents the radial part of the scalar field.

Substituting the metric tensor given in Eq.~(\ref{bb1}) together with the scalar field ansatz (\ref{sf2}) into the Klein-Gordon equation (\ref{sf1}), we obtain a Schr\"odinger-like wave equation of the form
\begin{equation}
\frac{d^2\psi}{dr_*^2} + \left[\omega^2 - V_{\text{s}}(r)\right]\psi = 0,
\label{sf3}
\end{equation}
where the tortoise coordinate $r_*$ is defined as
\begin{equation}
r_* = \int \frac{dr}{f(r)}.
\label{sf4}
\end{equation}
The effective potential for scalar perturbations is given by
\begin{equation}
V_{\text{s}}(r) = \left[\frac{\ell(\ell+1)}{r^2} + \frac{f'(r)}{r}\right] f(r).
\label{sf5}
\end{equation}

Substituting the explicit form of the metric function $f(r)$ and simplifying, the scalar perturbation potential becomes
\begin{equation}
V_{\text{s}}(r) =
\frac{\left(1 - \frac{2M}{r} + \eta e^{-\gamma}\frac{Q^2}{r^2} + \frac{\lambda}{r}\ln\!\frac{r}{|\lambda|}\right)}{r^2}
\left[\ell(\ell+1) + \frac{2M}{r} - 2\eta e^{-\gamma}\frac{Q^2}{r^2} + \frac{\lambda}{r}\left(1 - \ln\!\frac{r}{|\lambda|}\right)\right].
\label{sf6}
\end{equation}

\begin{itemize}
    \item For $\eta=+1$, corresponding to ModMax black hole with PFDM, the perturbations potential simplifies as
    \begin{equation}
        V_{\text{s}}(r) =\frac{\left(1 - \frac{2M}{r} +e^{-\gamma}\frac{Q^2}{r^2} + \frac{\lambda}{r}\ln\!\frac{r}{|\lambda|}\right)}{r^2}
\left[\ell(\ell+1) + \frac{2M}{r} - 2 e^{-\gamma}\frac{Q^2}{r^2} + \frac{\lambda}{r}\left(1 - \ln\!\frac{r}{|\lambda|}\right)\right].\label{sf6a}
    \end{equation}
    \item For $\eta=-1$, corresponding to phantom ModMax black hole with PFDM, the perturbations potential simplifies as
    \begin{equation}
        V_{\text{s}}(r) =\frac{\left(1 - \frac{2M}{r} - e^{-\gamma}\frac{Q^2}{r^2} + \frac{\lambda}{r}\ln\!\frac{r}{|\lambda|}\right)}{r^2}
\left[\ell(\ell+1) + \frac{2M}{r} + 2 e^{-\gamma}\frac{Q^2}{r^2} + \frac{\lambda}{r}\left(1 - \ln\!\frac{r}{|\lambda|}\right)\right].\label{sf6b}
    \end{equation}
\end{itemize}

\begin{figure}[htbp]
    \centering
    \includegraphics[width=\textwidth]{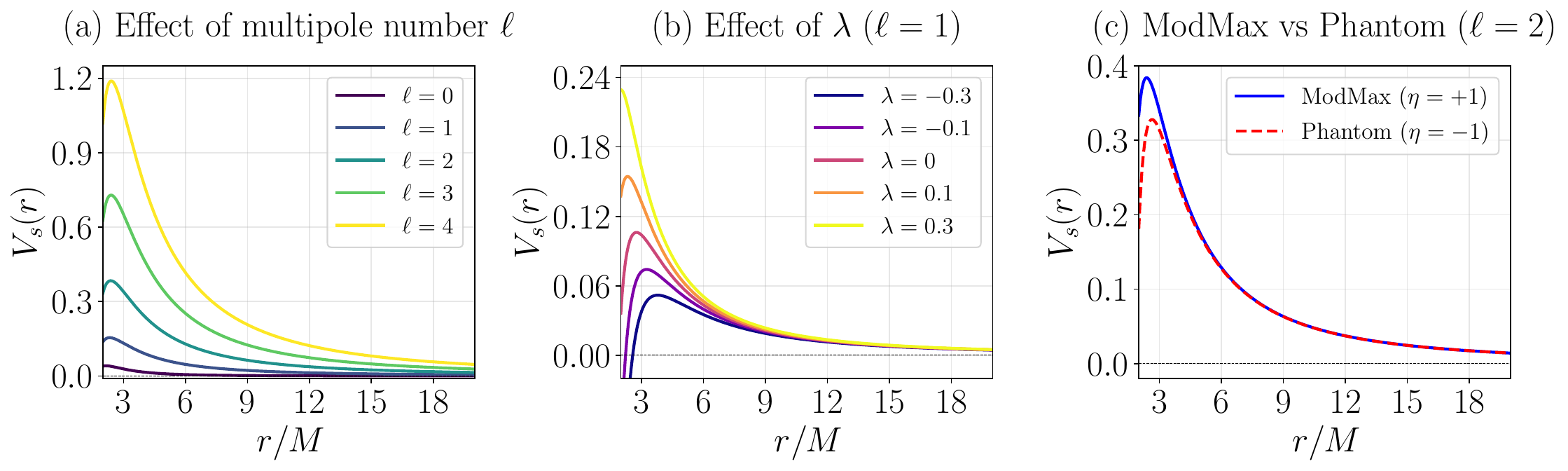}
    \caption{The effective potential $V_s(r)$ for scalar perturbations as a function of the radial coordinate $r/M$ in the Mod(A)Max black hole spacetime surrounded by perfect fluid dark matter. Panel~(a) demonstrates the effect of the multipole number $\ell$: higher $\ell$ values increase the height and shift the peak of the potential barrier outward, reflecting the centrifugal contribution $\ell(\ell+1)/r^2$. The potential approaches zero at large distances and vanishes at the horizon. Panel~(b) shows the influence of the PFDM parameter $\lambda$ for fixed $\ell = 1$; negative $\lambda$ values enhance the potential barrier and shift it to larger radii, while positive values suppress the barrier height. Panel~(c) compares ModMax ($\eta = +1$) and phantom ($\eta = -1$) configurations for $\ell = 2$, revealing that the phantom case exhibits a lower potential barrier located at larger radii. The height and shape of this potential barrier directly determine the quasinormal mode frequencies through the WKB approximation.}
    \label{fig:scalar_potential}
\end{figure}

The behavior of the scalar perturbation effective potential is illustrated in Fig.~\ref{fig:scalar_potential}, showing how the potential barrier depends on the multipole number $\ell$ and the model parameters.

The scalar perturbative potential derived above depends explicitly on the electric charge $Q$, the PFDM parameter $\lambda$, the ModMax parameter \(\gamma\), and the black hole mass \(M\). Moreover, the multipole number $\ell$ also alters this potential. Consequently, the quasinormal modes of the perturbation system also depend on these parameters.

Figure~\ref{fig:scalar_potential} provides the essential input for computing quasinormal modes via the WKB approximation. In the eikonal limit ($\ell \gg 1$), the QNM frequencies are determined by the properties of the potential at its maximum: the real part $\omega_{\Re}$ is related to the value of the potential at the peak, while the imaginary part $\omega_{\Im}$ (the damping rate) is determined by the curvature of the potential there.

The dependence on the multipole number in Fig.~\ref{fig:scalar_potential}(a) follows the expected $\ell(\ell+1)$ scaling from angular momentum conservation. Higher multipoles correspond to perturbations with more angular nodes, which are confined to smaller solid angles and thus experience stronger centrifugal repulsion.

The modification by the PFDM parameter shown in Fig.~\ref{fig:scalar_potential}(b) has direct observational consequences. A higher potential barrier (negative $\lambda$) leads to longer-lived QNMs with smaller imaginary parts, while a lower barrier (positive $\lambda$) produces more strongly damped oscillations. This provides a potential observational signature of dark matter effects in gravitational wave ringdown signals.

\begin{figure}[htbp]
    \centering
    \includegraphics[width=\textwidth]{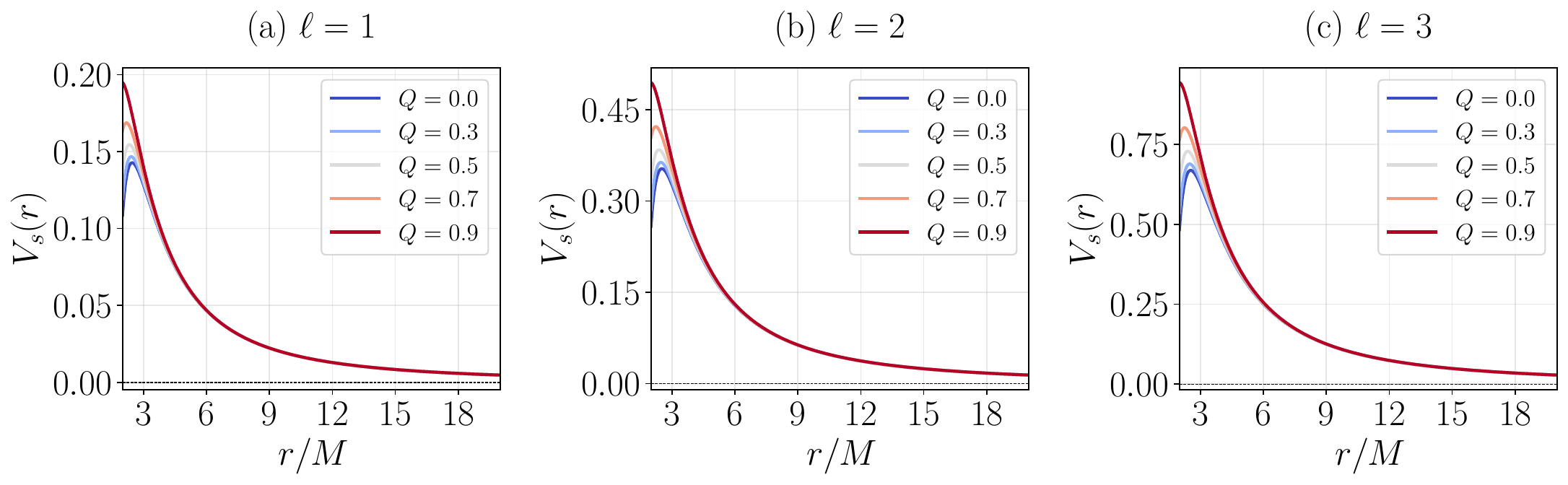}
    \caption{The effective potential $V_s(r)$ for scalar perturbations showing the combined effects of the electric charge $Q$ and multipole number $\ell$. Panel~(a) displays the potential for $\ell = 1$, panel~(b) for $\ell = 2$, and panel~(c) for $\ell = 3$. In each panel, curves from blue to red correspond to increasing charge values $Q = 0, 0.3, 0.5, 0.7, 0.9$ (in units of $M$). Increasing the charge raises the potential barrier and shifts its peak to smaller radii, consistent with the electromagnetic contribution to the effective geometry. The effect is more pronounced for lower multipoles, where the centrifugal term does not dominate. For high charges approaching extremality, the potential develops a double-peak structure near the degenerate horizon. The parameters are set to $\gamma = 0.3$, $\lambda = 0.1$, and $\eta = +1$ (ModMax).}
    \label{fig:scalar_potential_Q}
\end{figure}

The influence of the electric charge on the scalar potential is further explored in Fig.~\ref{fig:scalar_potential_Q}, which shows the potential for different charge values across multiple multipole modes.

The charge dependence shown in Fig.~\ref{fig:scalar_potential_Q} reveals important features of the QNM spectrum. As the charge increases, the potential barrier becomes higher and narrower, leading to higher frequency oscillations (larger $\omega_{\Re}$) but also stronger damping (larger $|\omega_{\Im}|$). Near extremality ($Q \to M$ for the Reissner-Nordstr\"om limit), the inner and outer horizons approach each other, and the potential develops characteristic features that signal the breakdown of the perturbative analysis.

\section{QNMs: Connection of QNMs with Shadow}

In the eikonal limit ($\ell \gg 1$), black hole (BH) quasinormal modes (QNMs) exhibit a well-known correspondence with the properties of the BH shadow. This relation is closely tied to the validity of the WKB approximation; once the WKB method ceases to be reliable, the correspondence may break down even in the large-$\ell$ regime \cite{SF1,SF2,SF3,SF4}. This remarkable connection was first established in \cite{VC2009}, where it was shown that the real part of the QNM frequencies is governed by the angular velocity $\Omega^{\rm null}_{\phi}(r_s)$ of unstable circular null geodesics, while the imaginary part is determined by the Lyapunov exponent $\lambda_L$, which characterizes the instability timescale of these orbits. Consequently, the QNM spectrum in the eikonal limit can be directly inferred from the properties of the photon sphere.

In this limit, the QNM frequencies are given by
\begin{equation}
\omega_{\ell \gg 1} = \Omega^{\rm null}_{\phi}(r_s)\, \ell - i \left(n + \frac{1}{2}\right) |\lambda_L|,
\label{ww1}
\end{equation}
where $n$ and $\ell$ denote the overtone number and the multipole number, respectively. The angular velocity of null geodesics is expressed as \cite{VC2009}
\begin{equation}
\Omega^{\rm null}_{\phi}(r) = \frac{d\phi}{dt}
= \frac{\dot{\phi}}{\dot{t}}
= \frac{\mathrm{L}}{r^2}\frac{f(r)}{\mathrm{E}}
= \frac{\sqrt{f(r)}}{r},
\label{ww2}
\end{equation}
where we have employed the circular null orbit condition given in Eq.~(\ref{bb8}).

Evaluating the above expression at the photon sphere radius $r=r_s$, we obtain
\begin{equation}
\Omega^{\rm null}_{\phi}(r_s) = \frac{\sqrt{f(r_s)}}{r_s}
= \frac{1}{R_{\rm sh}},
\label{ww3}
\end{equation}
where $R_{\rm sh}$ denotes the radius of the BH shadow given in (\ref{bb12}).

The Lyapunov exponent associated with the instability of null circular orbits is given by \cite{VC2009}
\begin{equation}
\lambda_L = \lim_{r \to r_s}
\sqrt{-\frac{1}{2}\frac{V''_{\rm eff}(r)}{\dot{t}^{\,2}}}
= \sqrt{\frac{f(r_s)\left[2f(r_s) - r_s^2 f''(r_s)\right]}{2r_s^2}}
= \Omega^{\rm null}_{\phi}(r_s)\sqrt{f(r_s) - \frac{r_s^2 f''(r_s)}{2}},
\label{ww4}
\end{equation}
which quantifies the decay rate of perturbations around the unstable photon orbit.

Using Eqs.~(\ref{ww1}) and (\ref{ww3}), the real part of the QNM frequency can be written as
\begin{equation}
\omega_{\Re} = \Omega^{\rm null}_{\phi}(r_s)\,\ell
= \lim_{\ell \gg 1} \frac{\ell}{R_{\rm sh}},
\label{ww5}
\end{equation}
demonstrating that larger shadow radii correspond to lower oscillation frequencies in the eikonal regime.

Furthermore, the Lyapunov exponent can be expressed directly in terms of the shadow radius as
\begin{equation}
\lambda_L = \frac{1}{R_{\rm sh}}\,\zeta,
\qquad
\zeta = \sqrt{f(r_s) - \frac{r_s^2 f''(r_s)}{2}}.
\label{ww6}
\end{equation}
Substituting this relation into Eq.~(\ref{ww1}), the QNM frequency in the eikonal limit becomes
\begin{equation}
\omega_{\ell \gg 1}
= \frac{1}{R_{\rm sh}}
\left[
\ell - i \left(n + \frac{1}{2}\right)|\zeta|
\right].
\label{ww7}
\end{equation}

\begin{figure}[htbp]
    \centering
    \includegraphics[width=\textwidth]{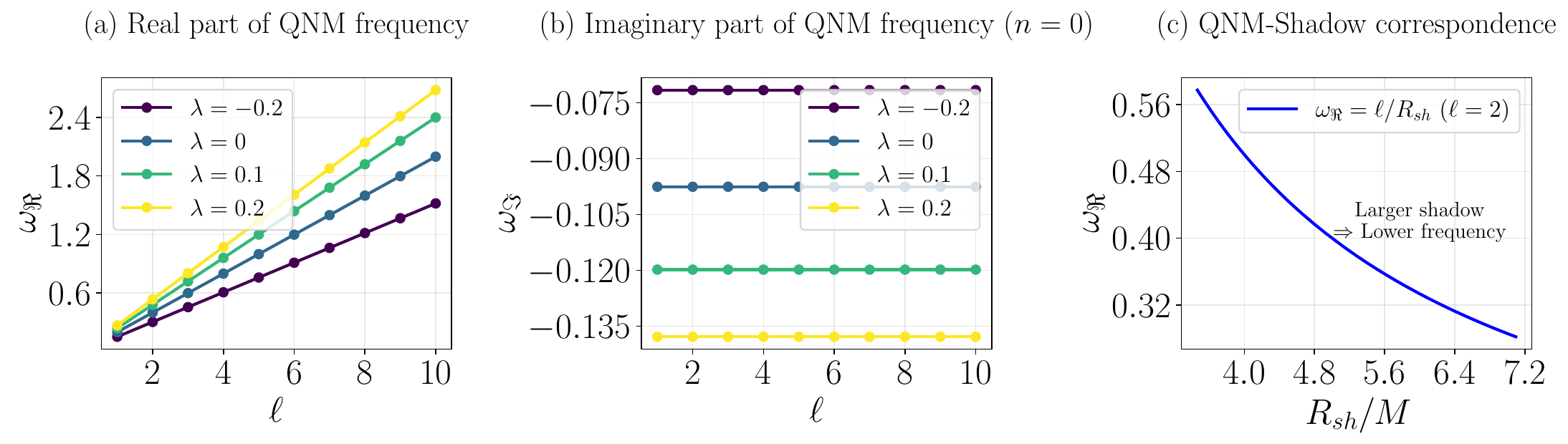}
    \caption{The connection between quasinormal modes and shadow properties for the Mod(A)Max black hole with PFDM. Panel~(a) shows the real part of the QNM frequency $\omega_{\Re}$ as a function of the multipole number $\ell$ for different values of the PFDM parameter $\lambda$. The linear relationship $\omega_{\Re} = \ell/R_{\rm sh}$ is evident, with the slope inversely proportional to the shadow radius. Negative $\lambda$ values (larger shadow) produce lower frequencies, while positive $\lambda$ values (smaller shadow) yield higher frequencies. Panel~(b) displays the imaginary part $\omega_{\Im}$ for the fundamental mode ($n = 0$), showing that the damping rate is independent of $\ell$ in the eikonal limit and is determined solely by the Lyapunov exponent. The PFDM parameter modifies the instability timescale of the photon orbit. Panel~(c) illustrates the direct relationship between the shadow radius $R_{\rm sh}$ and the QNM frequency for $\ell = 2$: larger shadows correspond to lower oscillation frequencies, providing an observational link between electromagnetic (shadow) and gravitational wave (ringdown) observations of black holes.}
    \label{fig:qnm_shadow}
\end{figure}

The QNM-shadow correspondence is illustrated in Fig.~\ref{fig:qnm_shadow}, which demonstrates how the observable properties of the black hole shadow directly constrain the gravitational wave ringdown spectrum.

The significance of this correspondence lies in the fact that the BH shadow radius can be directly inferred from astronomical observations. As a result, the real and imaginary parts of the QNM frequencies can be expressed in terms of observable shadow properties, offering an alternative approach to constraining BH parameters without explicitly relying on the geodesic method. This framework provides a complementary avenue for probing strong-field gravity through gravitational-wave observations of the ringdown phase.

Figure~\ref{fig:qnm_shadow} demonstrates the practical utility of the QNM-shadow correspondence. Figure~\ref{fig:qnm_shadow}(a) confirms the linear scaling predicted by Eq.~(\ref{ww5}), with the slope determined by the inverse shadow radius. The spread in slopes for different $\lambda$ values quantifies how PFDM affects the observable ringdown frequency. For the Event Horizon Telescope observations of M87* and Sgr~A*, the measured shadow sizes can be directly translated into predicted QNM frequencies using this relationship.

Figure~\ref{fig:qnm_shadow}(b) shows that the damping rate of QNMs in the eikonal limit is constant across multipoles, depending only on the Lyapunov exponent $\lambda_L$. This is a universal feature of the eikonal approximation and provides a clean prediction: measuring the damping time of the ringdown signal constrains the instability timescale of the photon orbit.

To complement the eikonal discussion with explicit numerical benchmarks, we now report representative QNM frequencies for massless scalar perturbations computed in the Mod(A)Max-PFDM geometry. Tables~\ref{taba1} and~\ref{taba2} summarize the dependence of the complex frequencies on the nonlinear electrodynamics parameter $\gamma$ at fixed PFDM strength $\lambda=1$ (with $n=0$, $M=1$, and $Q=1$). Comparing the ordinary ModMax branch ($\eta=+1$) with the phantom branch ($\eta=-1$) highlights a systematic shift in both the oscillation frequency and the damping rate: for the same $(\gamma,\ell)$, the phantom sector typically yields smaller $\Re(\omega)$ and a milder decay, consistent with a modified effective potential barrier and photon-sphere instability timescale.

\begin{table*}[!t]
\centering
\begin{tabular}{|c|c|c|c|}
\hline
\multicolumn{4}{|c|}{ Ordinary ModMax $\eta=1$ }\\
\hline
$\gamma$ & $\omega \left( \ell=1\right)$ & $\omega \left(\ell=2\right)$ & $\omega \left( \ell=3\right)$ \\
\hline
$0.2$ & $0.512676-0.604874i$ & $0.925321-0.542526i$ & $1.3168-0.523685i$ \\
$0.4$ & $0.485878-0.586657i$ & $0.879071-0.521252i$ & $1.25056-0.5016i$ \\
$0.6$ & $0.467207-0.572749i$ & $0.847285-0.505782i$ & $1.20521-0.485734i$ \\
$0.8$ & $0.453649-0.473989i$ & $0.824318-0.494235i$ & $1.172485-0.527126i$ \\
$1$   & $0.443515-0.4650925i$ & $0.807178-0.48544 i$ & $1.148086-0.563081i$ \\
$1.2$ & $0.435783-0.547114i$ & $0.783962-0.473308i$ & $1.12948-0.458236i$ \\
$1.4$ & $0.429792-0.541922i$ & $0.776009-0.469099i$ & $1.11506-0.452884i$ \\
$1.6$ & $0.425096-0.537787i$ & $0.776009-0.469099i$ & $1.10375-0.448664i$ \\
$1.8$ & $0.421381-0.534476i$ & $0.769715-0.465749i$ & $1.0948-0.445312i$ \\
$2$   & $0.418422-0.531814i$ & $0.764699-0.463068i$ & $1.08766-0.442632i$ \\
\hline
\end{tabular}
\caption{The QNM of ModMax BH with PFDM for a massless scalar perturbation with several values of $\gamma$. Here, $\lambda=1$, $n=0$, $M=1$, $Q=1$.}
\label{taba1}
\end{table*}

\begin{table*}[!t]
\centering
\begin{tabular}{|c|c|c|c|}
\hline
\multicolumn{4}{|c|}{ Mod(A)Max $\eta=-1$ }\\
\hline
$\gamma$ & $\omega \left( \ell=1\right)$ & $\omega \left(\ell=2\right)$ & $\omega \left( \ell=3\right)$ \\
\hline
$0.2$ & $0.350351-0.464852i$ & $0.648029-0.398432i$ & $0.921887-0.378793i$ \\
$0.4$ & $0.358541-0.473452i$ & $0.662221-0.406493i$ & $0.942045-0.386685i$ \\
$0.6$ & $0.36575-0.480906i$  & $0.674676-0.413528i$ & $0.959738-0.393587i$ \\
$0.8$ & $0.372034-0.487312i$ & $0.685503-0.419612i$ & $0.97511917-0.399567i$ \\
$1$   & $0.377463-0.492777i$ & $0.694835-0.424832i$ & $0.988377-0.404706i$ \\
$1.2$ & $0.382115-0.497408i$ & $0.702818-0.429277i$ & $0.999718-0.40909 i$ \\
$1.4$ & $0.386074-0.50131 i$ & $0.7096-0.43304 i$   & $1.00936-0.412806i$ \\
$1.6$ & $0.389423-0.504583i$ & $0.715329-0.436208i$ & $1.0175-0.415937i$ \\
$1.8$ & $0.39224-0.507317i$  & $0.720144-0.438863i$ & $1.02434-0.418564i$ \\
$2$   & $0.3946-0.509592i$   & $0.724174-0.44108 i$ & $1.03006-0.420759i$ \\
\hline
\end{tabular}
\caption{The QNM of Mod(A)Max BH with PFDM for a massless scalar perturbation with several values of $\gamma$. Here, $\lambda=1$, $n=0$, $M=1$, $Q=1$.}
\label{taba2}
\end{table*}

Next, we isolate the impact of the PFDM parameter $\lambda$ on the ringdown spectrum by fixing $\gamma=0.4$ and varying $\lambda$. Tables~\ref{taba3} and~\ref{taba4} display the corresponding QNMs for the ordinary ($\eta=+1$) and phantom ($\eta=-1$) branches, respectively, again for $n=0$, $M=1$, and $Q=1$. These data make explicit how increasing $\lambda$ lowers $\Re(\omega)$ while also shifting the damping rate, in agreement with the qualitative expectation from the QNM--shadow relation: stronger PFDM generally increases the characteristic length scale associated with the photon region (and hence the shadow), leading to smaller oscillation frequencies. The comparison between the two branches further indicates that the phantom sector responds differently to $\lambda$, providing an additional handle for phenomenological discrimination should both shadow and ringdown information become available for the same source.

\begin{table*}[!t]
\centering
\begin{tabular}{|c|c|c|c|}
\hline
\multicolumn{4}{|c|}{ Ordinary ModMax $\eta=1$ }\\
\hline
$\lambda$ & $\omega \left( \ell=1\right)$ & $\omega \left(\ell=2\right)$ & $\omega \left( \ell=3\right)$ \\
\hline
$1$   & $0.485878-0.586657i$ & $0.879071-0.521252i$ & $1.25056-0.5016i$ \\
$1.2$ & $0.470653-0.561148i$ & $0.842685-0.493667i$ & $1.19509-0.473199i$ \\
$1.4$ & $0.453384-0.534226i$ & $0.804704-0.466417i$ & $1.13834-0.445672i$ \\
$1.6$ & $0.435655-0.507803i$ & $0.767476-0.44072 i$ & $1.08338-0.420037i$ \\
$1.8$ & $0.418232-0.482711i$ & $0.732026-0.416966i$ & $1.0315-0.396554i$ \\
$2$   & $0.401486-0.459266i$ & $0.698747-0.395194i$ & $0.9831-0.375177i$ \\
\hline
\end{tabular}
\caption{The QNM of ModMax BH with PFDM for a massless scalar perturbation with several values of $\beta$. Here, $\gamma=0.4$, $n=0$, $M=1$, $Q=1$.}
\label{taba3}
\end{table*}

\begin{table*}[!t]
\centering
\begin{tabular}{|c|c|c|c|}
\hline
\multicolumn{4}{|c|}{ Mod(A)Max $\eta=-1$ }\\
\hline
$\lambda$ & $\omega \left( \ell=1\right)$ & $\omega \left(\ell=2\right)$ & $\omega \left( \ell=3\right)$ \\
\hline
$1$   & $0.358541-0.473452$ & $0.662221-0.406493$ & $0.942045-0.386685$ \\
$1.2$ & $0.36186-0.465487$  & $0.659639-0.398273$ & $0.935754-0.378141$ \\
$1.4$ & $0.360977-0.453934$ & $0.650671-0.387315$ & $0.920838-0.367145$ \\
$1.6$ & $0.357165-0.440408$ & $0.637628-0.374929$ & $0.900541-0.354918$ \\
$1.8$ & $0.351372-0.425949$ & $0.622112-0.361944$ & $0.877085-0.342223$ \\
$2$   & $0.344288-0.411215$ & $0.60522-0.348876$  & $0.851968-0.32953$  \\
\hline
\end{tabular}
\caption{The QNM of Mod(A)Max BH with PFDM for a massless scalar perturbation with several values of $\lambda$. Here, $\gamma=0.4$, $n=0$, $M=1$, $Q=1$.}
\label{taba4}
\end{table*}

The inverse relationship between shadow radius and QNM frequency shown in Fig.~\ref{fig:qnm_shadow}(c) has profound implications for multi-messenger astronomy. Joint observations of a black hole's shadow (via very long baseline interferometry) and its ringdown signal (via gravitational wave detectors) can be cross-validated using this correspondence. Any discrepancy would signal either a breakdown of general relativity or the presence of exotic matter fields---precisely the scenario explored in this work with PFDM.

The modification of this relationship by the PFDM parameter $\lambda$ provides a potential signature of dark matter in strong-field gravity. If the shadow size measured by the EHT is inconsistent with the QNM frequency inferred from gravitational wave observations (assuming standard general relativity), one possible explanation is the presence of a dark matter distribution around the black hole that modifies the geometry according to the model studied here.

\section{Conclusion}

We investigated the geodesic structure, optical signatures, perturbative dynamics, and thermodynamic behavior of a spherically symmetric Mod(A)Max black hole embedded in perfect fluid dark matter (PFDM), characterized by the parameters $(Q,\gamma,\lambda)$ and the branch choice $\eta=\pm1$. The analysis was organized around null and timelike geodesics, Hawking thermodynamics (including the Smarr relation), thermodynamic topology via the generalized Gibbs free energy, and massless scalar perturbations, with particular emphasis on the interplay between photon-sphere physics, shadow observables, and ringdown spectra.

On the optical side, the null-geodesic effective potential exhibits a single unstable barrier whose maximum defines the photon sphere. Because the lapse function contains the combined deformation $\eta e^{-\gamma}Q^2/r^2 + (\lambda/r)\ln(r/|\lambda|)$, both the photon-sphere radius $r_s$ and the shadow size $R_{\rm sh}$ respond sensitively to PFDM and nonlinear electrodynamics. Over the explored domain, negative $\lambda$ systematically enlarges the photon sphere and increases the shadow radius, whereas increasing $Q$ tends to reduce $R_{\rm sh}$, and increasing $\gamma$ suppresses the charge imprint through the factor $e^{-\gamma}$, driving the solution toward an effective Schwarzschild--PFDM behavior. The orbit integration further confirmed the standard three-regime classification (capture, near-critical spiraling, and scattering) and visualized the logarithmic sensitivity of the winding number of near-critical trajectories as $b\to b_c$, providing a direct geometric interpretation of higher-order photon rings.

For timelike motion, we derived the massive-particle effective potential and obtained closed expressions for the specific angular momentum and specific energy of circular orbits. The PFDM and Mod(A)Max parameters shift the location and depth of the potential minimum, thereby displacing the domain of stable circular motion and the innermost stable circular orbit (ISCO) inferred from the marginal stability condition. This displacement is physically relevant for accretion modeling because the ISCO location and the binding energy $1-\mathcal{E}_{\rm sp}$ set leading expectations for radiative efficiency and characteristic disk radii. We also showed that flipping the electromagnetic branch $\eta=+1\to -1$ (ordinary to phantom sector) produces systematic outward shifts of characteristic radii (horizon, photon sphere, and potential minima), consistent with the sign reversal of the electromagnetic contribution in $f(r)$.

Thermodynamically, we obtained analytic expressions for the mass $M(r_h)$, Hawking temperature $T(r_h)$, Bekenstein--Hawking entropy $S=\pi r_h^2$, and heat capacity $C(r_h)$. The parameter dependence of $T$ and $C$ reveals regimes of local instability ($C<0$) and critical points where $C$ diverges, signaling second-order phase transitions in the canonical ensemble. Treating $\lambda$ as a thermodynamic variable, we derived the conjugate potential $\Phi_\lambda$ and verified the Smarr-like relation \(M = 2TS + \Phi Q + \Phi_\lambda \lambda,\) thereby confirming the internal consistency of the extended first law for the Mod(A)Max-PFDM system.

To characterize the \emph{global} phase structure, we adopted the thermodynamic-topology correspondence based on the generalized Gibbs free energy $\mathcal{F}$ and its associated vector field. The zeros of the vector field identify on-shell equilibrium configurations and act as topological defects in the auxiliary $(r_h,\theta)$ space; their winding numbers encode stability information in a coordinate-independent way. The resulting zero-point curve $\tau=1/T$ compactly encodes the accessible equilibrium branches, and its deformation with $(Q,\gamma,\lambda,\eta)$ clarifies how PFDM and nonlinear electrodynamics reshape the phase portrait beyond what is visible from local quantities alone.

We also analyzed massless scalar perturbations, deriving the Schr\"odinger-like master equation and the effective potential $V_s(r)$. The potential barrier depends on $\ell$ and on $(Q,\gamma,\lambda,\eta)$ through $f(r)$ and $f'(r)$, implying that the quasinormal mode (QNM) spectrum carries a complementary imprint of PFDM and Mod(A)Max electrodynamics. In the eikonal regime, we highlighted the QNM--shadow correspondence, where the real part of the QNM frequency scales as $\omega_{\Re}\simeq \ell/R_{\rm sh}$ and the damping is governed by the Lyapunov exponent of the unstable null orbit. The numerical benchmarks reported for several $(\gamma,\lambda)$ values are consistent with this geometric picture and further indicate that the phantom branch can produce systematically different oscillation and decay rates than the ordinary branch, suggesting a potential phenomenological discriminator in multi-messenger settings.

Overall, our results provide a unified framework in which (i) optical observables (photon sphere and shadow), (ii) orbital structure (timelike effective potential and ISCO), (iii) thermodynamic response (temperature, heat capacity, and topology), and (iv) dynamical stability diagnostics (scalar potentials and QNMs) are all controlled by the same deformation parameters $(Q,\gamma,\lambda)$ and by the branch choice $\eta=\pm1$. This creates a natural pathway for confronting Mod(A)Max--PFDM phenomenology with future observations: shadow measurements constrain $R_{\rm sh}$ and hence the photon-sphere sector, while ringdown data constrain the associated instability scale through QNMs, and thermodynamic/topological diagnostics clarify which parameter ranges support stable equilibrium branches.

As a natural continuation, it would be interesting to (a) extend the construction to rotating geometries to assess degeneracies between spin and PFDM/Mod(A)Max deformations in shadow and QNM observables, (b) compute QNMs beyond the eikonal approximation (and for other perturbation spins) to test the robustness of the shadow--ringdown correspondence in this background, and (c) incorporate additional observables (lensing, time delays, or disk spectra) to break parameter degeneracies between $(Q,\gamma,\lambda)$ and the branch choice $\eta$.

\footnotesize

\section*{Acknowledgments}

F.A. acknowledges the Inter University Centre for Astronomy and Astrophysics (IUCAA), Pune, India for granting visiting associateship.

\section*{Solution of Cubic Equation}

\[
r^3-6Mr^2+9\eta e^{-\gamma}Q^2r
-\frac{4\eta^2 e^{-2\gamma}Q^4}{M}=0 .
\]

Introduce the shift
\[
r=x+2M ,
\]
which removes the quadratic term and yields the depressed cubic
\[
x^3+px+q=0 ,
\]
with
\[
p=9\eta e^{-\gamma}Q^2-12M^2 ,\qquad 
q=18M\eta e^{-\gamma}Q^2-16M^3
-\frac{4\eta^2 e^{-2\gamma}Q^4}{M}.
\]

Define the discriminant
\[
\Delta=\left(\frac{q}{2}\right)^2+\left(\frac{p}{3}\right)^3 .
\]

A real solution is given by Cardano’s formula
\[
x=
\sqrt[3]{-\frac{q}{2}+\sqrt{\Delta}}
+
\sqrt[3]{-\frac{q}{2}-\sqrt{\Delta}} .
\]

Transforming back to $r$, the solution reads
\[
\boxed{
r=2M
+\sqrt[3]{-\frac{q}{2}+\sqrt{\Delta}}
+\sqrt[3]{-\frac{q}{2}-\sqrt{\Delta}}
}
\]

\end{document}